\documentclass[12pt]{iopart}
\usepackage{comment}
\usepackage{graphicx}
\newcommand{\zlabel}[1]{\label{#1} }
\newcommand{\fc}{\frac} 
\newcommand{\rt}{\right} 
\newcommand{\mrr}{\mathbf{r}}
\newcommand{\pr}{\prime}

\newcommand{\mx}{\mathbf{x}}

\newcommand{\al}{\alpha}

\newcommand{\pa}[2]{\frac{\partial #1}{\partial #2}}

\newcommand{\bv}{\mathbf{v}}
\newcommand{\bu}{\mathbf{u}}
\newcommand{\bw}{\mathbf{w}}

\newcommand{\dive}{\text{\bf div}\;\!}
\newcommand{\grad}{\text{\bf grad}\;\!}

\newcommand{\lab}[1]{ (\ref{#1}):\; }
\newcommand{\labb}[1]{ (\ref{#1})\; }
\newcommand{\dens}{\Upsilon}

\newcommand{\veps}{\varepsilon}
\newcommand{\sPsi}{\mbox{\small $\Psi$}}
\newcommand{\defi}{\,\,\dot{=}\,\,}
\newcommand{\bE}{E}
\newcommand{\Eqline}[1]{\vspace{1ex}\noindent\centerline{$#1$}\vspace{1ex}}
\newcommand{\mathbb}{\mathbf}
\newcommand{\lvert}{\vert}
\newcommand{\rvert}{\vert}
\newcommand{\text}{\mbox}
\newcommand{\boldsymbol}{\mathbf}
\newcommand{\alt}[1]{}

\begin{document}

\title[Fields and Equations of Classical Mechanics for Quantum Mechanics]{Fields and Equations of Classical Mechanics for Quantum Mechanics}

\author{James P. Finley}

\address{Department of Physical Sciences,
Eastern New~Mexico University,
Portales, NM 88130}
\ead{james.finley@enmu.edu}

\date{\today}

\begin{abstract}
A generalized Euler-equation of fluid dynamics is derived for describing many-body states of
quantum mechanics for fermion systems. The derived Eulerian equation can be viewed as representing the
interaction of two substates, where each substate has its own velocity and pressure fields.
These field quantities are given by maps depending on the probability distribution and the
phase of the wavefunction.
For one-body systems, the Eulerian equation can model either a fluid or particle
interpretations for quantum-mechanical states, where the streamlines of the Madelung, or
probability, fluid are also the trajectories of the particles.
%
%
For the fluid interpretation, the mass density is the probability density times the electron
mass.  The generalized Euler equation is shown to be the gradient of an equation representing
the total-energy of the two substates, having two energy fields that are, in general,
nonuniform. This total-energy equation is a generalization of the Bernoulli equation of fluid
dynamics. The total-energy equation, along with a continuity-equation,
is equivalent to the time-dependent Schr\"odinger equation.
An equation is also derived that is equivalent to the main equation of Bohmian mechanics.
%
%
However, additional identifications are given that are not part of Bohmian mechanics: The
quantum potential of Bohmian mechanics is given as a sum of a kinetic energy and pressure
fields. Also, the time derivative of the wavefunction phase is replaced by an energy
field.
%
In the formalism, field quantities are identified from their placement in equations of
classical mechanics.
Separately, the field quantities are given by definitions that involve the wavefunction and operators of quantum mechanics.
This approach
yields, unintended, and unknown energy and pressure fields. These fields, however, are shown to
satisfy a continuity equation, an equation that is equivalent to the other equation of Bohmian
mechanics.
It is also demonstrated that energy conservation holds for both of these energy fields, if the
wavefunction is a linear-combination of eigenvectors, where the eigenvectors  can be nondegenerate.
A detailed investigation is given on the possible behavior, or source, of an electron that has
one of the velocity fields. Alternate formulae for this velocity fields are also considered.
\end{abstract}

\maketitle



\section{Introduction}
\alt{The introduction and abstract is rewritten.}

There is a large spectrum of models used to describe physical phenomena.  The metric used to
evaluate a model, from this spectrum, depends on the academic field where the model is employed.
In this paper, quantum-mechanical states are modeled using concepts from classical
mechanics. In order to determine the best metric to evaluate this type of model, it is useful
to determine where such a model fits into the spectrum of models.

Starting, at one end of the spectrum, is the many-worlds interpretation of quantum
mechanics. In this interpretation, there are an infinite number of universes. As far as
philosophy is concerned, the impact that such a model has in science is immaterial.  Instead,
it is evaluated as a deterministic model that does not use a collapse theory to treat the
measurement problem.  As a useful scientific tool, this model would need to provide verifiable
predictions or, at least, provide physical scientists a useful way to think about
quantum-mechanical states.

Another model is the De Broglie--Bohm theory
\cite{Bohm:52a,Bohm:52b,B2,B4,B5,B7,B8,B9,Jung,Renziehausenb}, which we call Bohmian
mechanics. As a philosophical theory, it is deterministic with hidden variables.  In the
spectrum of models, it is in the same neighborhood as the many-worlds interpretation.  In
Bohmian mechanics, in the treatment of an electronic system, each electron, at all times, has
assigned to it a definite position and momentum. The trajectory of all particles are determined
by an initial configuration, a set of momentums and positions. Since---short of a scientific
revolution---it is not possible to determine the particles positions and momentums
simultaneously, for all practical purposes, the method is indeterminate. The questions of being
deterministic being purely philosophical.

On the other, far end of the spectrum, is the theory of Lewis dot structures used in
chemistry. In this model, dots are used to represent valence electrons of atoms and molecules,
with rules determining where the dots are placed. The Lewis dot structures contain information
about regions of high electron-density, and the electron density is considered a continuum,
called a charge cloud. In a typical organic chemistry textbook, there are thousands of Lewis
dot structure, but none can be found in physical chemistry or molecular physics textbooks.
Rigorous quantum-mechanical methods have not displaced the use of this model. Instead, both
models coexists.  For example, in crystal field theory \cite{Zumdahl}, lone-pair
valence-electrons of ligands split the degeneracy of the d orbitals of the transition-metal
cation of a transition metal complex, providing a means to predict if the complex will absorb
light in the visible spectrum.

In more rigorous methodology, the charge clouds become the probability densities of electrons
from either molecular-orbital, valence-bond theory, or other ab initio approaches
\cite{Szabo,Parr:89}. Being a static model, the electron cloud interpretation is, however,
incomplete, having only electrostatic fields.  An assignment of a kinetic-energy field, where
the change cloud becomes a non-static fluid of charged-mass, could have many applications, and
even improve, or supplement, the Lewis dot structure theory.  Such a kinetic-energy field could
be used in conjunction with the electrostatic external potential-energy and the
electron-electron repulsion-energy, both being classical fields that are widely used in
chemistry.

Another method on the spectrum is quantum hydrodynamics \cite{B6,Takabayasi}. This method is
closer to the electron-cloud interpretation than the many worlds interpretation, even though it
uses equations and ideas from Bohmian mechanics
\cite{Bohm:52a,Bohm:52b,B2,B4,B5,B7,B8,B9,Jung,Renziehausenb}. For one-body states, this method
also incorporates ideas from the Madelung fluid \cite{Madelung:26,Madelung:27}, where
Bohmian-mechanics and Madelung-mechanics overlap, sharing the same set of equations and
velocity field.

One literal way to view the model of hydrodynamics is as a fluid, where the mass density is
the probability density (times the electron mass), and the Bohmian velocity is the velocity
field. One way to introduce the thinking behind this identification is to consider the
following sequence of conclusions: 1) Since is it not known which Bohmian trajectories to use,
all trajectories are used. 2) Since none of the trajectory lines cross, it is useful to
identify the set of all such lines as the streamlines of a fluid-velocity field. 3) Using this
Madelung fluid, which is called a ``probability fluid,'' the model can be applied to the study
of quantum mechanical states. It this methodology it is in immaterial if there is such a fluid.
Instead, the information contained in the fluid can be viewed as a way to organize information
from the wavefunction.  (Note that the Madelung formalism has been generalized to treat
many-body systems \cite{Renziehausenb}.)

Compared to the information assigned to fluids of classical mechanics, quantum hydrodynamics
assigns less information, containing only a velocity field and a mass-density. This suggests
that it is incomplete, that there are more classical analogs to be identified.  Also, the
Hamilton-Jacobi equation of Bohmian mechanics, also used in hydrodynamics, contains the quantum
potential $Q$ \cite{Bohm:52a,Bohm:52b,B6}, which has no classical analog. The dependence of the
quantum potential $Q$ on the Laplacian, of the wavefunction, suggests that the velocity field is
missing kinetic energy, since kinetic-energy is ``stored'' in the quantum potential.  Also the
probability fluid is not assigned a pressure, a field present in all classical fluids.

In hydrodynamics, the velocity field is determined by the phase of the wavefunction. Therefore,
the probability fluid of hydrodynamics is not the charge cloud of ab initio theories, since
the charge cloud is determined by the probability density. This suggests that these two approaches
can be combined into one model.

The assignments of classical-mechanical analogs to functions and equations derived from quantum
mechanics have useful applications, as in the many applications of quantum hydrodynamics,
discussed below. It is also useful to have core sets of mathematical relationships that are
used for different disciplines, as with the Laplace equation. Two mathematical systems, based
on completely different axioms, can be equivalent. With this in mind, various models, based on
very different identifications, can coexists, with no controversy. Such models require a
different metric to evaluate their effectiveness than the metrics used in philosophical theory.

In this field of classical identification, the identification of a classical physical-property
is made by a mathematical analogy.  As a generalization of the electron cloud identification,
various forms of energy are defined over all space, such that regions of space contain
quantitative information about the different energy forms.  Satisfactory assignments, of
classical mechanical fields to functions from quantum mechanics, should have some
correspondence with the quantum states they are applied to.  For example, a
classical-mechanical velocity field, to be a good match, should have some correspondence with
the expectation value of the kinetic energy.

Recently \cite{Finley-Arxiv,Finley-Bern}, progress has been made in this area of research, with
the derivation of an equation that is identified as one from classical mechanics. This equation
contains field quantities of classical mechanics, given as formalae of the probability density,
applicable to a class of quantum states: The time-independent Schr\"odinger equation for
one-body stationary states with real-valued wavefunctions was shown to be equivalent to a
compressible-flow generalization of the Bernoulli equation of fluid dynamics.  The
kinetic-energy, pressure and mass-density fields are identified from their presence in the
Bernoullian equation.  The kinetic-energy field naturally yields to the identification of
velocity and momentum fields.
The derived generalized Bernoulli equation describes compressible, irrotational, steady flow
with \emph{local} variable mass. Over all space, mass is conserved, because the rate of mass
creation from the sources are equal to the rate of mass annihilation from the sinks. Also, each
fluid element has a constant energy per mass.

This work is continued in this paper, where both the quantum hydrodynamics and the
Bernoullian-fluid models are naturally combined into one general model, applicable to all
many-body states of quantum mechanics for fermion systems.
A total-energy equation is derived. This equation is a generalization of the equations from
quantum hydrodynamics and the Bernoullian equation.  The total-energy equation can be viewed as
a sum of two equations, corresponding to two interacting systems: One of the systems has
variable mass from the Bernoullian fluid, and the other one has conserved mass from the
Madelung, or probability, fluid. Also, each of the two systems have their own velocity,
pressure and energy fields. A generalized Euler equation of fluid dynamics is shown to be
implied by the total-energy equation.

Physical properties of classical mechanics are identified as field quantities appearing in
equations of classical mechanics, and these fields are maps of the probability distribution and
wavefunction phase.  The two velocity fields from the Beroullian fluid and the Madelung fluid
are identified in this way. A pressure is also identified as the one from the Bernoullian
equation. Furthermore, an energy field is identified as a nonuniform-field generalization of
the eigenvalue of the time-independent Schr\"odinger equation.

A second, independent, method is also presented for the identification of physical properties
of classical mechanics for quantum-mechanical systems. These physical properties are also given as
fields determined by the wavefunction.
For example, two classical momentum-fields, given as an ordered pair, are defined, in a natural
way, using the momentum operator and wavefunction of quantum mechanics. (The ordered pair is
the real and imaginary components of the integrand of the expectation-value of the momentum
operator, divided by the probability distribution.)
%
This definitions yields the same two corresponding velocity fields as the ones mentioned in the
previous paragraph, where they were identified by their appearance in classical-mechanic
equations.  In a similar way, order pairs of field quantities are defined for the energy and
pressure. This approach yields the same formulae for the energy and pressure fields mentioned in
the previous paragraph.

This approach also gives two unintended consequences: Additional, unknown pressure and energy
fields. However, these two field are found to be present in the total-energy equation. They are
also shown to appear in an equation that is equivalent to the continuity equation of Bohmian
mechanics.

These last two, unintended field quantities, together with the other identified field
quantities, gives a classical-mechanical representation, of quantum mechanical systems, that is
complete, as far as energy is concerned, where each and every term from the total-energy
equation has a classical analog. In the special case of continuity-equation satisfaction, the
total-energy equation splits into two equations, that, together with the formulae for the field
quantities, are equivalent to the time-dependent Schr\"odinger equation. Hence, the methodology
is also ab initio.

Also, the two nonuniform energy-fields are shown to be conserved in a case where the
wavefunction of the time-dependent Schr\"odinger equation is not necessarily an eigenfunction
of the Hamiltonian operator.

While the treatment of energy is complete, the identification of the source of the kinetic
energy corresponding to the velocity identified from the Bernoullian equation is an open
questions. Progress is made on this issued by  exploring of many possibilities.  In an
attempt to exhaust all possibilities, both particle and fluid descriptions are considered.

Before moving to a description of what is done in the following sections, the next paragraph
gives a brief overview of other theoretical results. Applications from quantum hydrodynamics and
related fields are also mentioned.

Heifetz and coworkers \cite{Heifetz,Heifetz2} explores the thermodynamics of Madelung fluids.
%
There are many generalizations of the Madelung equations \cite{Sorokin,Broadbridge,Schonberg,Caliari,Jamali}.
The generalization by Broadbridge \cite{Broadbridge} and Jamali \cite{Jamali} use a complex velocity.
Tsekov \cite{Tsekov} also uses a complex velocity to derives a complex Navier--Stokes equation.
Vadasz \cite{Vadasz} derived an extension of the Schr\"odinger equation from the Navier--Stokes
equation. Quantum hydrodynamic theory has been employed to treat systems with single particle wave functions
\cite{6,7,8,9,10,11,12,14,15,16,17,18,19,20,21,22}.
The method also also been generalized to treat many particle systems
\cite{8b,Renziehausen}. Application of this formalism include the investigation of spin effects
\cite{14b,15b}, Bose--Einstein condensates \cite{16b}, graphene \cite{17b} and plasmas
\cite{18b,22b,23b}.

Paragraphs that follow indicate the sequence of derivations and results from this paper. Some
of the notation used in this overview are introduced. To reduce clutter, this notation is for
the special case of one-body expressed in atomic units. Section~\ref{4925} gives a
self-contained overview that is easier to follow than the explanations in this
introduction. This occurs, because that presentation is not restricted to a sequence that
follows the order that the results are obtained. Some readers may prefer to read that material,
instead, or in addition to, what follows. Readers can also skip directly to the sections with
results, where a brief introduction is given of what is accomplished.

The background needed to fully comprehend the material in this paper involves elementary quantum-
and fluid-mechanics. This material is covered in the beginning chapters of many
monographs, including Levine \cite{Levine}, for quantum mechanics, and Munson, Young, and
Okiishi \cite{Munson}, for fluid mechanics.

Section~\ref{p7835} demonstrates that a many-body generalization of the Bernoulli equation of
classical fluid-dynamics is equivalent to the time-independent Schr\"odinger equation
($\hat{H}\Psi = E\Psi$), in the special case where the wavefunction is real valued.
The given derivation is a generalization of a
derivation for the special case of one-body \cite{Finley-Arxiv,Finley-Bern}.  The
kinetic-energy $mu^2/2$ and pressure $P$ fields, applicable to quantum mechanical systems, are
identified from their presence in the Bernoullian equation:
($mu^2/2 + P/\rho + U = E$),
an equation of energy~$E$ conservation, involving the external potential $U$.  The
kinetic-energy field $mu^2/2$ naturally yields an identification of a velocity field $\bu$,
and a corresponding particle momentum $m\bu$. As far as energy is concerned, there is some
flexibility in the choice of the velocity direction, especially the sign. This flexibility in sign
is indicated by the notation $\bu_\pm$, such that ($\bu_- = -\bu_+$).


Each one of the identified fields of Sec.~\ref{p7835} is defined by a formula involving the
probability distribution $\rho$ and/or derivatives of the probability distribution, e.g., ($P =
-\nabla^2\rho/4$).  It is understood that the equivalence of the Schr\"odinger and Bernoullian
equation, mentioned above, requires these mathematical definitions, so, strictly speaking, the
equivalence involves an equation set.  The derived Bernoullian equation reduces to the well
known Bernoulli equation of fluid dynamics for the case of one-body.
%

Section~\ref{4592} gives interpretations, in terms of particles, of the fields of the
Bernoullian equation, derived in Sec.~\ref{p7835}, for quantum states that satisfy this
equation. The kinetic-energy $mu^2/2$ and pressure $P$ fields are then shown to be related to
the expectation-value $\langle\Psi\vert\mbox{$-\nabla^2$}/2\vert\Psi\rangle$ of the kinetic
energy: The kinetic-energy integrand $\Psi\mbox{$(-\nabla^2/2)$}\Psi$ is equal to the sum of
the kinetic- and pressure-energy fields (per volume), i.e., $\rho_mu^2/2 + P$, where
($\rho_m = m\vert\Psi\vert^2$).  The pressure field is also shown to vanish when integrated over
all space, implying that the kinetic-energy field $\rho_mu^2/2$ can replace the kinetic-energy
integrand $\Psi\mbox{$(-\nabla^2)$}/2\Psi$ in the calculation of the expectation-value of the
kinetic energy. This result gives an additional, and independent, classical identification of
both the kinetic-energy $mu^2/2$ and velocity $\bu$ fields  that agrees with the assignments
given in Sec.~\ref{p7835}.

Section~\ref{p4242} works on the main equation from Bohmian mechanics, where the wavefunction
is written in polar form ($\Psi = Re^{iS}$), defined by functions ($R^2 = \rho$) and $S$.  The
identification from Sec.~\ref{p7835} involving the kinetic energy, mentioned above, is used to
identify the well-known quantum potential $Q$ from Bohmian mechanics, as a sum of two fields,
involving the kinetic energy and pressure, i.e, ($Q = mu^2/2 + P/\rho_m$). When this sum
replaces the quantum potential in the Hamilton-Jacobi equation of Bohmian mechanics---that
already contains one kinetic-energy field $mv^2/2$---the resulting Hamilton-Jacobi equation
becomes
($mv^2/2 + mu^2/2 + P/\rho + U = -\partial S$),
where $\partial S$ is the time derivative of the phase, and the equation now has two kinetic
energy fields, $mv^2/2$ and $mu^2/2$, and one pressure field $P$. The two kinetic energy fields
yield two velocity fields, $\bv$ and $\bu$, and two corresponding particle momentums, $m\bv$
and $m\bu$: One momentum $m\bv$ is from Bohmian mechanics, and the other one $m\bu$ is from the
Bernoullian equation developed in Sec.~\ref{p7835}.

The resulting Hamilton-Jacobi equation, mentioned in the previous paragraph, together with the
continuity equation, is demonstrated to be equivalent to the time-independent Schr\"odinger
equation.  For the special case of stationary states, this equation reduces to a generalization
of the Bernoullian equation developed in Sec.~\ref{p7835}, holding also for complex valued
wavefunctions. Also, for stationary states, we have ($\bE = -\partial S$), where $\bE$ is the
energy eigenvalue $\bE$ of the Schr\"odinger equation, a uniform field.

For the remainder of Sec.~\ref{p4242}, it is demonstrated that if the two velocity fields are
orthogonal, i.e., ($\bv\cdot\bu = \mathbf{0}$), then the continuity equation of Bohmian
mechanics, ($\partial\rho = -\nabla\cdot\rho\bv$), reduces to a generalization of the Poisson
equation, and to a Laplace equation for stationary states. This material in not part of the
main logical sequence, and it is not needed for the later results.

In quantum mechanics, momentum, an observable, is defined via axioms involving the momentum
operator $\hat{P}$. In contrast, the above mentioned momentum-particle definitions come from
identifications in equations that are implied by Schr\"odiger equations. It is, therefore,
reasonable to investigate if the momentum operator itself can naturally give alternate
definitions for the same particle momentums mentioned above.  In Sec.~\ref{2370}, this approach
is investigated, yielding two particle-moment definitions, as a pair of fields, defined by a
complex-valued function $\Psi^*\hat{P}\Psi/\rho$, involving the momentum operator
$\hat{P}$. One of the particle momentums is taken as the real part of the function, and the
other momentum is taken as the imaginary part. The resulting pair of momentums, called
momentae, $m\bv$ and $m\bu$, are the same ones defined in Secs.~\ref{p7835} and \ref{p4242}.
Furthermore, a natural definition for a kinetic energy-field is also given, giving the same
field as in Secs.~\ref{4592} and \ref{p4242}. Since the momentums fields, $m\bv$ and $m\bu$,
are irrotational, they are also expressed by their potentials: $S$ and ($\theta =
-\hbar\ln\rho$), respectively.

Sec.~\ref{5147} defines a notational system that reduces clutter in equations for
many-body systems, and yields, in most case, many-body equations that are displayed
exactly the same way as in the special case of one-body. In order that this system is not
confused with sloppy notation, where subscripts are suppressed and understood, the
presentation is formal.

After the diversion of Sec.~\ref{5147}, the main logical sequence is continued in
Sec.~\ref{2371}, where, as in the particle momentae definitions in Sec.~\ref{2370}, two energy
fields are defined as the real and imaginary parts of the function $\Phi^*\hat{H}\Psi/\rho$,
where the $\hat{H}$ is the Hamiltonian (energy) operator. Using the Schr\"odinger
Eq.~($i\Phi^*\partial\Psi = \Phi^*\hat{H}\Psi$), the approach yields a pair of energy fields as
the real and imaginary parts of $i\Phi^*\partial\Psi/\rho$: ($\bE _S = -\partial S$) and ($\bE
_\theta = -\partial\theta$), where $S$ and $\theta$ are the momentae potentials, mentioned
above.

One of the energy fields $\bE_S$, identified by the development of Sec.~\ref{2371}, assigns an
energy field to any solution of the time-dependent Schr\"odinger equation.  When ($-\partial S =
\bE _S $) is substituted into the Hamilton-Jacobi equation developed in Sec.~\ref{p4242}, this
equation becomes ($\bE_S = mv^2/2 + um^2/2 + P/\rho + U$), an energy equation.
Furthermore, the field $E_S$ reduces to the energy eigenvalue $E$ of the Schr\"odinger
equation in the case of time independence.

At this point in Sec.~\ref{2371}, the other energy field $\bE_\theta$ is an unknown, unintended
consequence of the development. However, later in Sec.~(\ref{4379}), this field appears in an
energy form of the continuity equation, an equation equivalent to one of the equations of
Bohmian mechanics, implied by the Schr\"odinger equation.

Sec.~(\ref{2475}) extends the method of definition for energies and momentae fields to
pressures, where, unlike the other cases, the pressure is not an observable of quantum
mechanics. A definition is given, such that the two pressures are proportional to the real and
imaginary parts of the divergence of the momentum (per volume),
$\nabla\cdot(\Psi^*\hat{P}\Psi)$. This definitions yields the same pressure field $P$ mentioned
above.  The second pressure $P_\bv$, an unknown, appears in the same continuity equation as the
energy field $E_\theta$, mentioned above. (The notation for the other pressure is changed from
$P$ to $P_\bu$.)

Sec.~\ref{4379} uses the continuity equation to derive a number of relations involving the
pressures. The energy equation ($\bE_\theta = P_\bv/\rho$), mentioned in the previous
paragraph, is obtained in the development, containing both the unknown pressure $P_\bv$ and
energy $\bE_\theta$ fields. This energy equation is shown to be equivalent to the continuity
equation. Hence, the three defined ordered pairs---$\{m\bv,m\bu\}$, $\{P_\bv,P_\bu\}$, and
$\{\bE_S,\bE_\theta\}$---are accounted for, appearing in two equations that, together, are
equivalent to the time-dependent Schr\"odinger equation. Each and every term is identified from
the two equations, one term $U$ being the sum of the classical external-potential and
electron-electron repulsion-energy.

Sec.~\ref{p1830} combines the two energy equations, giving a total-energy equation
($\bE  =  \fc{1}{2}m v^2 + P_\bv\rho^{-1} + \fc12 m u^2 + P_\bu\rho^{-1} + U$),
where ($\bE \defi \bE _S + \bE _\theta$).  A many-body generalization of the Euler equation of
fluid dynamics is derived from this equation. For one-body states, this Eulerian equation can
be viewed as a sum of two equations, corresponding to two interacting systems. One with
variable mass with velocity $\bu$ and pressure $P_\bu$; one with conserved mass with velocity
$\bv$ and pressure $P_\bv$. The \emph{local} time derivative of the velocity fields are shown
to satisfy ($-\nabla\bE _S = m\partial\bv$) and ($-\nabla\bE _\theta = m\partial\bu$), where,
for the case of one spin-free body, values $\partial\bv(\mrr,t)$ and $\partial\bu(\mrr,t)$ are
the time rate of change of the velocity fields at \emph{fixed} position $\mrr\in\mathbb{R}^3$
and time $t$. (These are not accelerations of fluid elements.)

Sec.~\ref{4290} investigates the important property of energy conservation for the two energy
fields, $\bE_\theta$ and $\bE_s$, where these fields are not, in general, uniform. The
wavefunctions considered are linear combinations of functions, where each function is an
eigenfunction of the same Hamiltonian operator. If such a wavefunction contains nondegenerate
eigenfunctions, it does not satisfy the time independent Schr\"odinger equation.  It is
demonstrated that, in this case, the two energy quantities are conserved over all space for all
times: The two forms of energy can flow in space, but neither energy is created or destroyed.

Sec.~\ref{5382} considers various issues involving velocity. Sec.~\ref{4382} investigates a
velocity compatibility problem for the two velocity fields, where the vector sum $\bv + \bu$
does not give the correct total kinetic-energy, if $\bv\cdot\bu\ne \mathbf{0}$. Sec.~\ref{1025}
considers other possibilities for the velocity $\bu$ from the Bernoullian equation, especially
the velocity direction.  From non-relativistic quantum mechanics, a velocity field is derived
that has the same direction as a well known one, as an approximation from the Pauli equation.
Various issues of local variable mass, angular momentum, and kinetic energy-satisfaction are
considered, for both fluid and particle descriptions. While progress is made, the question of
the best choice of Bernoullian velocity $\bu$ remains open.

\section{A Classical Mechanics Energy Equation for Stationary States with Real Valued Wavefunctions of Quantum Mechanics \zlabel{p7835}}

In this section, a many--body generalization of the Bernoulli equation of fluid-dynamics is
shown to be equivalent to the time-independent Schr\"odinger equation, in the case where the
wavefunction is real-valued.  The kinetic-energy $mu_i^2/2$ and pressure $P_i$ fields,
applicable to quantum-mechanical systems, are identified from their presence in the Bernoullian
equation, and these fields have formulae dependent on the probability distribution $\Upsilon =
\vert\Psi\vert^2$.  The kinetic-energy fields $mu_i^2/2$ naturally yields the identification of
a $n$ velocity fields $\bu_i$ and particle momentums $m\bu_i$.  The main result of this section
is the Bernoullian Eq.~(\ref{p4791b}), that reduces to (\ref{p2574}), for the case of one body
with probability distribution ($\rho = \Upsilon$).

The $n$-body time-independent Schr\"odinger equation with a normalized, real-valued eigenfunction $R$,
can be written
\begin{equation} \zlabel{p2602}
  -\fc{\hbar^2}{2m}\sum_{i=1}^n R\nabla_i^2R + \sum_{i=1}^n V_i\Upsilon + \fc12 \sum_{i\ne j}^n W_{ij}\Upsilon = \bE \Upsilon,
\end{equation}
where 
\[
\left[R\nabla_i^2R\right](\mx) = R(\mx)\nabla_{\mrr_i}^2R(\mx), \quad \mx = \mathbf{x}_1,\cdots \mathbf{x}_n.
\]
Also, the $n$-body probability distribution $\Upsilon$ is $\Upsilon = R^2$; the electron
coordinate $\mathbf{x}_i$ is defined by $\mathbf{x}_i = \mathbf{r}_i,\omega_i$, where
$\mathbf{r}_i\in\mathbb{R}^3$ and $\omega_i\in\{-1,1\}$ are the spatial and spin coordinates,
respectively. Furthermore, the functions $V_i$ and $W_{ij}$ are spin-free multiplicative
operators. For $n$-electron systems, these are given by the following:
\begin{equation} \zlabel{7152}
[V_i\Upsilon](\mx) =  V(\mrr_i)\Upsilon(\mx), \quad [W_{ij}\Upsilon](\mx) = \fc{e^2}{4\pi\varepsilon_0}\vert\mrr_i - \mrr_j\vert^{-1}\Upsilon(\mx),
\end{equation}
where $-e$ is the electron charge, $\varepsilon_0$ is the permittivity constant, and the one-body
external potential $V$ is a specified real-valued function with domain $\mathbb{R}^3$ such that
$\{\mx\vert R(\mx) = 0\}$ has measure zero. This requirements for $V$ implies that the division
of an equation by $R$ or $\Upsilon$ gives an equation that is defined almost everywhere (a.e.).

\alt{P 10/L 2: Removed reference to appendix, removed appendix, and modified the sentence accordingly.}

Substituting the following equality
\begin{equation} \zlabel{0152A}
-\fc12\left[R\nabla_i^2R\right] = \fc18\left[\Upsilon^{-1}\nabla_i\Upsilon\cdot\nabla_i\Upsilon\right] - \fc14\nabla_i^2\Upsilon,
\end{equation}
that is proved elsewhere \cite{Finley-Arxiv,Finley-Bern}, into the Schr\"odinger equation (\ref{p2602}), we obtain
\[
\fc{\hbar^2}{8m}\sum_{i=1}^n\Upsilon^{-1} \left\vert\nabla_i\Upsilon\right\vert^2- \fc{\hbar^2}{4m}\sum_{i=1}^n\nabla_i^2\Upsilon  +
  \sum_{i=1}^n V_i\Upsilon + \fc12 \sum_{i\ne j}^n W_{ij}\Upsilon = \bE \Upsilon. 
  \]

\alt{P 10/ L 19: ``Defined the term ``Bernoulli velocity.''}
  
Using the mathematical definitions
\begin{equation} \zlabel{p4720}
\bu_{i\pm} = \pm\fc{\hbar}{2m}\fc{\nabla_i\Upsilon}{\Upsilon},
\end{equation}
\begin{equation} \zlabel{p4722}
  P_i = -\fc{\hbar^2}{4m}\nabla_i^2\Upsilon,
\end{equation}
where we call $\bu_{i\pm}$ the Bernoullian velocity, and
\begin{equation} \zlabel{2047}
\fc12 m u_i^2 = \fc12 m\bu_{i\pm}\cdot\bu_{i\pm} = \fc12 m \left\vert\fc{\hbar}{2m}\fc{\nabla_i\Upsilon}{\Upsilon}\right\vert^2
= \fc{\hbar^2}{8m} \Upsilon^{-2} \left\vert \nabla_i\Upsilon\right\vert^2,
\end{equation}
we have
\begin{equation} \zlabel{p4791}
\sum_i^n \fc12 \Upsilon m u_i^2  + \sum_i^n P_i +   \sum_{i=1}^n V_i\Upsilon + \fc12 \sum_{i\ne j}^nW_{ij}\Upsilon = \bE \Upsilon.
\end{equation}
For any point $\mx$ such that $\Upsilon(\mx) \ne 0$, this energy equation can also be written
\begin{equation} \zlabel{p4791b}
  \sum_i \left(\fc12 m u_i^2  + P_i\Upsilon^{-1}\right) + U = \bE ,
\end{equation}
where
\begin{equation} \zlabel{p0025}
U = \fc12 \sum_{i\ne j}^n W_{ij} +  \sum_{i=1}^n V_i.
\end{equation}
Since the derivation of equation (\ref{p4791b}) from the Schr\"odinger equation (\ref{p2602}) is
reversible, Eqs.~(\ref{p2602}) and (\ref{p4791b}) are equivalent, i.e., $\Upsilon$ is a solution
of (\ref{p4791b}) a.e, if and only if $\Upsilon$ is a solution of (\ref{p2602}) a.e.

Next consider a state of a one-body system, such that $R(\mrr,\pm 1) = \phi(\mrr)\alpha(\pm 1)$, where
$\alpha$ is the spin function that satisfies $\alpha(1)= 1$ and $\alpha(-1)= 0$. Hence,
$\Upsilon(\mrr,1) = \phi^2(\mrr)\;\dot{=}\;\rho(\mrr)$, and the last equality defines the one-body
probability density~$\rho$.  In this special case, with $U = V$, (\ref{p4720}), (\ref{p4722}) and
(\ref{p4791b}) can be written
\begin{equation}
\zlabel{p2572}  \bu_{\pm} = \pm\fc{\hbar}{2m}\fc{\nabla\rho}{\rho}, \quad P = -\fc{\hbar^2}{4m}\nabla^2\rho,
\end{equation}  
\begin{equation}
\zlabel{p2574}  \fc12 m u^2 + P\rho^{-1} + U = \bE.
\end{equation}
\alt{ P 11/L 9 Added sentence to the end of the paragraph: ``The fields $m\bu_\pm$ and
  $\rho_m\bu_\pm$ are called particle- and fluid-momentums, respectively.'' Clarified a point
  about the energies $\bE \rho$ and $\bE/m$ of the fluid elements.}

These equations have been used to treat one-body stationary-states of quantum-mechanical
systems with real-valued wave-functions as flows of a fluid, where Eq.~(\ref{p2574}) is
identified as a compressible-flow generalization of the Berrnoulli equation with body force
$U$, pressure $p$, velocity field $\bu_\pm$, and mass density~$\rho_m = m\rho$
\cite{Finley-Arxiv,Finley-Bern}. Also, $\bE \rho$ is the energy per volume, and $\bE/m$ is the
energy per mass, of the fluid elements.  Furthermore, $m$ is total the mass of the fluid, equal
to the electron mass. The fields $m\bu_\pm$ and $\rho_m\bu_\pm$ are called particle- and
fluid-momentums, respectively.

\alt{Added paragraph to discuss other work from the literature.}

The velocity choice $\bu_\pm$ of Eq.~(\ref{p2572}) for one-body systems, appears in other
investigations.  The function $\varrho_m\vert\bu_{\pm}\vert^2/2$ is a term of the Hamiltonian
functional of the generalized fluid-dynamics formalism by Broer \cite{Broer}, where the
Hamiltonian functional is derived from the time-dependent Schr\"odinger equation, and
$\varrho_m$ satisfies $\varrho_m = 2\rho_m$.  Salesi \cite{Salesi} obtains a Lagrangian
function that is equivalent to the Madelung equations, and this function contains the term
$\rho_m\vert\bu_{\pm}\vert^2/2$. He shows that a variational approach of the Lagrangian has
$\rho_m\vert\bu_{\pm}\vert^2/2$ determining the quantum potential $Q$ of Bohmian mechanics.  He
interprets $\vert\bu_{\pm}\vert^2/2$ as the internal energy of the relative motion in the
center of mass coordinate frame from the \mbox{Zitterbewegung} (ZWB) model of spin. The
velocity magnitude $\vert\bu_\pm\vert$ follows as a non-relativistic approximation of a
velocity expression of Hestenes \cite{Hestenes} of Schr\"odinger--Pauli theories.  Furthermore,
Tsekov \cite{Tsekov} obtains the same velocity choice $\bu_\pm$ of Eq.~(\ref{p2572}) as the
imaginary component of a complex velocity, where the formalism involves diffusion.

The velocity choice $\bu_-$, called downhill flow, has the fluid particles move in the
direction of lower density; similarly, choice $\bu_+$ is called uphill flow.  For later use, we
note that the Bernoullian velocity definition (\ref{p4720}) can be generalized to
\begin{equation} \zlabel{p4720x}
  \bu_{i\pm} = u_{i\pm}\mathbf{\hat{s}}_i,
  \quad u_{i\pm}  = \pm\fc{\hbar}{2m}\fc{\left\lvert\nabla_i\Upsilon\right\rvert}{\Upsilon},
\end{equation}
where $\mathbf{\hat{s}}_i = \mathbf{\hat{s}}_i(\mrr)$ is a unit vector at our disposal. This
generalizations yields kinetic energies $mu_i^2$ that are the same as (\ref{2047}), and the
energy equation (\ref{p4791b}) does not depend on the unit vector $\mathbf{\hat{s}}_i$.

\section{Interpretations Based on Classical Mechanics \zlabel{4592}}

As in Bohmian Mechanics \cite{Bohm:52a,Bohm:52b}, in this section, we present a particle
interpretation for one-body states that satisfy the energy energy~(\ref{p2574}), and then
the interpretation is generalized to many-body states that satisfy the Bernoullian
Eq.~(\ref{p4791b}).
The kinetic-energy integrand of quantum mechanics is shown to be equal to the sum of the
kinetic- and pressure-energy fields (per volume), as indicated in Eq.~(\ref{p2922}).
The pressure fields are shown to vanish when integrated over all space, implying that the
kinetic-energy field $\rho_m(u_1^2\, + \cdots u_n^2)/2$ can replace the kinetic-energy
integrand in the calculation of the expectation-value of the kinetic energy. This result gives
an additional, and independent, classical identification of both the kinetic-energy $mu_i^2/2$
and velocity $\bu_i$ fields that agrees with the assignments given in Sec.~\ref{p7835}.

For one-body states, when the particle is at position $\mrr\in\mathbb{R}^3$, we interpret
$\lvert\bu_\pm\rvert(\mrr)$ and $P(\mrr)$ as the speed and pressure, respectively, defined by
Eqs.~(\ref{p2572}).  Also, for the energy equation~(\ref{p2574}), we take $[mu^2/2]$,
$[p\rho^{-1}]$, $U$, and $\bE$ to be the kinetic-, ``compression''-, potential-, and
total-energy of the particle, respectively, where these scalar fields have domain
$\mathbb{R}^3$, and $\bE$ is uniform.
Since the kinetic energy depends on the particle's mass and speed only, we only require the
velocity to satisfy Eq.~(\ref{p4720x}). Thus, we do not require the velocity direction to be
either of the directions given by irrotational vector field (\ref{p2572}). Further
consideration of the velocity choice is given in Sec.~(\ref{1025}), including a well known
velocity field that is derived from relativistic quantum-mechanics.

The interpretations for one-body states have a natural generalization to many-body states, with
energy equation (\ref{p4791b}).  For example, as in Bohmian mechanics, the configuration
$\mx = \mathbf{x}_1,\cdots \mathbf{x}_n$ is taken to mean that one electron is located at
$\mrr_1$ with spin-variable value $\omega_1$, another one is at $\mrr_2$ with $\omega_2$, and
so on.  The function
\begin{equation} \zlabel{5322}
  \fc12 m [u_i(\mx_1,\mx_2,\cdots \mx_i,\cdots \mx_n)]^2,
\end{equation}
is interpreted as the kinetic energy of the $i$th particle from configuration $\mx$, that is,
it is the kinetic energy of a particle located at $\mx_i$, where the $i$th particle speed is
$\lvert u_{i\pm}(\mx)\rvert$.
Similar interpretations are given to the other energy terms from (\ref{p4791b}), and $P_i(\mx)$
can be interpreted as the pressure subjected to the $i$th particle from configuration~$\mx$.
With these interpretations, except for the spin dependence and the probability distribution
$\dens$, equation~(\ref{p4791b}) is interpreted as a classical energy equation $H = \bE$,
with a Hamiltonian function $H$ that depends on the probability distribution $\dens$, partial
derivatives of the probability distribution, and the potential energy function~$U$.

\alt{P 12/ L 23: Reference to Eq.~(8) changed to Eq.~(7).}

By comparing (\ref{p2602}) and (\ref{p4791}), we obtain an equality satisfied by the
\emph{integrand} of the expectation-value of the kinetic energy from quantum mechanics:
\begin{equation} \zlabel{p2922}
-\fc{\hbar^2}{2m}\sum_{i=1}^n [R\nabla_i^2R] = \dens \sum_i^n\left(\fc12 m u_i^2 + P_i\dens^{-1}\right).
\end{equation} 
From this equation and the above particle interpretations, it follows that the value at the
point $\mx$ of the \emph{integrand}, of the expectation value of the kinetic energy, is the sum
of the $n$ particle kinetic- and compression-energies for configuration $\mx$, multiplied by
the weight $\dens$, and $\dens$ is required to be normalized, i.e.,
\[
\sum_{\omega_1 \cdots\omega_n} \int_{\mathbb{R}^{3n}} \dens  = 1.
\]
Note that the compression energy $P_i\Upsilon^{-1}$ times the probability distribution $\Upsilon$ is
the pressure $P_i$ for the $i$th particle.

\alt{Clarified the meaning of $P_{\mathbf{x}_i^\pr}(\mx_i)$.}

For each Cartesian coordinate $\al_i\in\{x_i,y_i,z_i\}$, ($i = 1,\cdots n$), we require the
wavefunction to satisfy
\[
\lim_{\al_i \to \pm\infty} R(\mx) = \lim_{\al_i \to \pm\infty} \pa{R}{\al_i} = 0.
\]
Hence
\[
\int_{-\infty}^{\infty} \pa{^2\dens}{\al_i^2}\, d\al_i = \left.\pa{\dens}{\al_i}\right\rvert_{-\infty}^{\infty} =
2\left.R\pa{R}{\al_i}\rt\rvert_{-\infty}^{\infty} = 0,
\]
and therefore
\[
\int_{\mathbb{R}^3} \nabla_i^2\dens\, d\mrr_i = 0.
\]
This result combined with (\ref{p4722}) gives
\begin{equation} \zlabel{p5280}
\int_{\mathbb{R}^3} P_i(\mx)\, d\mrr_i = 0.
\end{equation}
Therefore, it follows from Eq.~(\ref{p5280}), that for a fixed ``partial'' configuration
\[
\mathbf{x}_i^\pr = \mx_1\cdots\mx_{i-1}, \mx_{i+1}\cdots\mx_n,
\]
where $\mx_i$ is excluded, that the function 
\[
P_{\mathbf{x}_i^\pr}(\mx_i) = P_i(\mx_1\cdots\mx_m), \quad \mx_i\in\mathbb{R}^3,
\]
where $\mathbf{x}_i^\pr$ is considered a parameter,  must have both positive and negative
values on subspaces with nonzero measures, or be the zero function a.e. Also, it follows from
Eqs.~(\ref{p2922}) and (\ref{p5280}) that the pressure $P_i$ does not contribute to the
expectation value of the kinetic energy.

\alt{P13/L 42: In the last sentence of the paragraph ``where the brackets $\langle\cdots\rangle$
  indicate an expectation value'' is change to ``where $\langle T\rangle[\dens]$ is the expectation value
  of the kinetic energy.''}

Using equality (\ref{p5280}) for $P_i$, and integrating (\ref{p2922}) over the $3n$ spatial
coordinates and summing over the $n$ spin coordinates from the set $\Omega$, we have
\begin{equation} \zlabel{p4321}
\langle T\rangle[R] \;\dot{=}\; -\fc{\hbar^2}{2m}\sum_{\Omega}\int_{\mathbb{R}^{3n}}\sum_{i=1}^n R \nabla_i^2 R
= \sum_{\Omega}\int_{\mathbb{R}^{3n}}\sum_{i=1}^n\fc12\Upsilon\hspace{.2ex} mu_i^2.
\end{equation}
Hence 
\begin{equation} \zlabel{4024}
\langle T\rangle[\dens] =  \left\langle\sum_{\Omega}\sum_{i=1}^n\fc12 mu_i^2 \right\rangle,
\end{equation}
where $\langle T\rangle[\dens]$ is the expectation value of the kinetic energy.
This result supports the interpretations given in Sec.~\ref{p7835} for the kinetic energy
and speed.  For a one-body system represented by the spin orbital $\phi\al$, as is done in the
derivation of Eqs.~(\ref{p2572}) and (\ref{p2574}), Eq.~(\ref{4024}) reduces to
\[
\langle T\rangle = \int \fc12 \rho_m  u^2\, d\mrr,
\]
where $\rho_m = m\rho$, and for a fluid interpretation, $\rho_m$ is the mass density.



\section{Extended Bohmian Mechanics \zlabel{p4242}} 

\alt{A introductory paragraph.}

In this section, the kinetic-energy integrand result~(\ref{p2922}) from the previous section is
used to identify the quantum potential $Q$ from Bohmian mechanics. This relation is substituted
into the Hamilton-Jacobi equation of Bohmian mechanics. The gives result~(\ref{p5522}),
containing the pressures $P_i$ and velocity $\bu_i$ fields from the Bernoullian
equation~(\ref{p4791b}), and also the original velocity $\bv_i$, defined by Eq.~(\ref{p3312})
from Bohminan Mechanics. Also, $S$ is the wavefunction phase of the polar
form~($\Psi=Re^{iS/\hbar}$) of the wavefunction. The resulting equation~(\ref{p5522}), together
with the continuity equation~(\ref{p4288}) of Bohminan mechanics, along with the given
definitions for the fields, are shown to be equivalent to the time-dependent Schr\"odinger
equation~(\ref{p2588}). For the special case of stationary states, we have ($-\partial S = E$),
where $E$ is the eigenvalue of the time independent Schr\"odinger equation~(\ref{p2588}), and
the resulting equation~(\ref{p9922}) is a generalization of the Bernoullian Eq.~(\ref{p4791b}),
holding also for complex valued wavefunction.

\alt{P 15/L 3: Added a sentence before the last sentence, stating that Equations (20) and (21),
  taken together are equivalent to Eq..~(18).}

The time-dependent Schr\"odinger equation is \cite{Levine,Bransden}
\begin{equation} \zlabel{p2588}
i\hbar\partial\Psi = -\fc{\hbar^2}{2m}\sum_i^n\nabla_i^2\Psi + U\Psi = \hat{H}\Psi,
\end{equation}
where $\partial\Psi(t) = \partial\Psi/\partial t$, $U = U(\mx,t)$ is given by (\ref{p0025}),
$\Psi = \Psi(\mathbf{x},t)$ is the $n$-body time-dependent wavefunction, and we use the same
notation as in the previous sections, e.g., $\mx_i = \mrr_i,\omega_i$.  Let the spin coordinates
$\omega_i,\cdots \omega_n$ be specified parameters. Hence, $\Psi = \Psi(\mrr,t)$, where $\mrr =
\mrr_1,\cdots \mrr_n$, permitting us to consider $\Psi$ to be a function of the time and the
spatial coordinates only. For Bohmian mechanics
\cite{Bohm:52a,Bohm:52b,B2,B4,B5,B6,B7,B8,B9,Jung,Renziehausenb}, the wavefunction is
represented in polar form is
\begin{equation} \zlabel{p4225}
\Psi = Re^{iS/\hbar},
\end{equation}
where $R$ and $S$ are time-dependent real-valued functions. When this ansatz is substituted into the
time-dependent Schr\"odinger equation (\ref{p2588}), after significant manipulations,
the following two equations can be obtained \cite{Renziehausenb}:
\begin{equation} \zlabel{p4288}
  \partial \Upsilon + \sum_{i=1}^n\nabla_i\cdot(\Upsilon\bv_i)  = \mathbf{0},
 \end{equation}
\begin{equation} \zlabel{p4299} 
  -R \partial S = \sum_{i=1}^n\left(\fc{1}{2m}R \nabla_i S\cdot\nabla_i S - \fc{\hbar^2}{2m}\nabla_i^2 R  \right) + UR,
\end{equation}
where the probability distribution is $\Upsilon = \Psi\Psi^* = R^2$, $\partial S(\mx,t)=
\partial S(\mx,t)/\partial t$, and the velocity $\bv_i$ of the $i$th particle is defined below.
Equation~(\ref{p4288}) is called the continuity equation.
In the special case of a one-body system, with $\Upsilon =\rho$, this equation has the same
form as the continuity equation from fluid dynamics \cite{Munson,Shapiro}, a statement of the
conservation of mass, where the mass density is $m\rho$.  The above two equations (\ref{p4288})
and (\ref{p4299}) are identical to Eqs.~(12) and (6) in the manuscript by Renziehausen and
Barth \cite{Renziehausenb}, for the special case considered here where there is only one kind
of particle, e.g., only electrons. Also, by examining the mathematics used in the derivation of
(\ref{p4288}) and (\ref{p4299}), it is easy to demonstrate that these two equations, taken
together, are equivalent to the Schr\"odinger Eq.~(\ref{p2588}).  The quantum potential $Q$,
presented below, is defined by Eq.~(18) in the same manuscript.

\alt{P 15/L 14: After Eq.~(\ref{p3322}), defined de Broglie velocity and later used the definition.}

Bohmian mechanics also defines the following two functions:
\begin{equation}\zlabel{p3312}
  \mathbf{v}_i = \fc{\nabla_iS}{m} = \text{Im}\left(\fc{\hbar}{m}\fc{\nabla_i\Psi}{\Psi}\right),
\end{equation}
\begin{equation} \zlabel{p3322}
  Q = -\fc{1}{R}\fc{\hbar^2}{2m}\nabla_i^2 R = \Upsilon^{-1}\left(-R\fc{\hbar^2}{2m}\nabla_i^2 R\right),
\end{equation}
where $Q$ is known as the Bohm quantum potential \cite{Bohm:52a,Bohm:52b,B6}, and we call
$\mathbf{v}_i$ the de Broglie velocity.  The function value $\mathbf{v}_i(\mx_1,\cdots \mx_i,\cdots
\mx_n) $ is interpreted as the velocity of the $i$th particle, i.e., the velocity of the
particle located at $\mx_i$ for the configuration $\mx = \mathbf{x}_1,\cdots \mathbf{x}_n$.
The identity given in (\ref{p3312}) is included because sometimes the second definition is used
when discussing Bohmian mechanics.  This identity is proved in Appendix~\ref{p7722}, where a
relation between the two velocities, $\bv_i$ and $\bu_i$, is also given.

\alt{P15/L 29. Two sentences at the end added to demonstrate that Eqs.~(24) and (20) are equivalent to the
Schr\"odinger Eq.~(18).}

Substituting the above two definitions into (\ref{p4299}), and dividing by $R$, we get
\begin{equation} \zlabel{p5200}
 -\partial S = \sum_i\fc{1}{2}m v_i^2  + Q  + U,
\end{equation}
where $v_i^2 = \vert\mathbf{v}_i\vert^2$. This is a Hamilton--Jocobi equation \cite{B6} with
the addition of the quantum potential $Q$. By direct substitution, it is easily seen that
Eq.~(\ref{p5200}) along with definitions~(\ref{p3312}) and (\ref{p3322}) is equivalent to
Eq.~(\ref{p4299}). Hence, Eq.~(\ref{p5200}), together with definitions~(\ref{p3312}) and
(\ref{p3322}), and the continuity equation~(\ref{p4288}), are equivalent to the Schr\"odinger
Eq.~(\ref{p2588}).

\alt{P 15/L 45 Added at the end, work by another author is mentioned, supporting the quantum potential interpretation.}

Note that Eq.~(\ref{p2922}) is an equality holding for two times differentiable real-valued
functions, where $u_i^2 = \vert\bu_{i\pm}\vert^2$ and $P_i$ are given be Eq.~(\ref{p4720}) and
(\ref{p4722}), respectively.
Next we extend the interpretations of the functions $\bu_{i\pm}$ and $P_i$ from Eqs.~(\ref{p4720})
and (\ref{p4722}) to the case where $R$ is the real-valued factor of the time-dependent wavefunctions
$\Psi$, given by ansatz~(\ref{p4225}), and note that these functions also appear in
(\ref{p2922}).
Making these interpretations and substituting Eq.~(\ref{p2922}) into (\ref{p3322}), we discover
\begin{equation} \zlabel{p5202}
Q = \sum_i\fc12 m u_i^2 + \sum_iP_i\Upsilon^{-1},
\end{equation}
where $Q$ is a sum of the $n$-particle kinetic energy $\sum_i\fc12 m u_i^2$ and the
$n$-particle compression-energy $\sum_iP_i\Upsilon^{-1}$. The kinetic energy portion of this
quantum-potential $Q$ expression agrees with the one from Salesi \cite{Salesi}, where no
interpretation is given for the other term of $Q$.

\alt{P 16/L 11: removed ``Eqs.~(\ref{p4288}) and (\ref{p5522}) are equivalent to the
  Schr\"odinger equation~(\ref{p2588}).'' This information is moved and expended.}

Substituting (\ref{p5202}) into (\ref{p5200}) gives the desired result:
\begin{equation} \zlabel{p5522}
-\partial S = \sum_i\left(\fc{1}{2}m v_i^2  + \fc12 m u_i^2 + P_i\Upsilon^{-1}\right) + U.
\end{equation}
This equation is a further development of (\ref{p5200}), containing two kinetic energy terms, a
compression energy term $\sum_i P_i\Upsilon^{-1}$, and the external potential $U$, given by
(\ref{p0025}). It seems reasonable at this point to assume that the total velocity of the $i$th particle is
$\bu_{i\pm} + \mathbf{v}_i$.  
Eq.~(\ref{p5522}) is a variant of the Hamilton--Jocobi equation.
The right-hand-side of (\ref{p5522}) can be interpreted as the time-dependent energy,
i.e., a Hamiltonian function. For the left-hand side, from Eq.~(\ref{p3312}), $S$ can be
interpreted as the momentum potential for each of the $n$ particles, but only including
the de Broglie velocity $\mathbf{v}_i$ portion of the total velocity.

\alt{P 16/L 20 Added a 2 sentence paragraph to demonstrate equivalence of Eq.~(29), together with Eq.~(20)
the Schr\"odinger Eq.~(18).}

By direct substitution, Eqs.~(\ref{p5200}) and (\ref{p5202}) are equivalent to
Eq.~(\ref{p5522}).  Hence, Eq.~(\ref{p5522}) together with Eq.~(\ref{p4288}, along with the
above definitions for the fields $v_i$, $u_i$, and $P_i$, are equivalent to the Schr\"odinger
Eq.~(\ref{p2588}).

\alt{P 16/L 30 Starting at ``Multiplying'', derived a generalization of a previous result~(\ref{4024}).}

If $\Psi$ is a stationary state then $\Psi(\mx) = R(\mx)e^{-i\bE t/\hbar}$, giving $S(t) = -\bE
t$. Hence, from (\ref{p5522}), we have 
\begin{equation} \zlabel{p9922}
\sum_i\left(\fc{1}{2}m v_i^2 + \fc12 m u_i^2 + P_i\Upsilon^{-1}\right) + U = \bE .
\end{equation}
This equation is a generalization of Eq.~(\ref{p4791b}), holding for complex valued
wavefunctions.
Multiplying this equation by $\Upsilon$, using the time-independent Schr\"odinger equation,
($\Psi^*\bar{H}\Psi = \bE\Upsilon$) with Hamiltonian definition~(\ref{p2588}), integrating
the result over the $3n$ spatial coordinates, summing over the $n$ spin coordinates, and
using also Eq.~(\ref{p5280}), we find that
\begin{equation*} 
 \langle T\rangle[\dens] =  \left\langle\sum_{\Omega}\sum_{i=1}^n\left(\fc12 mu_i^2 + \fc12 mv_i^2\right)\right\rangle, 
\end{equation*}
a generalization of Eq.~(\ref{4024}). This result supports the interpretation of speed and kinetic
energy involving the velocity vector $\bv_i$.

Note that the Bernoullian velocity $\bu_{i\pm}$ does not solve a continuity equation like
(\ref{p4288}), instead, from (\ref{p4720}) and (\ref{p4722}), we have
\begin{equation} \zlabel{p0254}
\nabla_i\cdot(\bu_{i\pm}\Upsilon) = \pm\fc{\hbar}{2m}\nabla_i^2\Upsilon
= \mp\fc{2}{\hbar}P_i.
\end{equation}
This equation should not be confused with the continuity equation implied by the time-dependent
Schr\"odinger equation that implies probability conservation. That equation remains satisfied,
since our probability distributions under consideration have wavefunctions that satisfy the
time dependent Schr\"odinger equation.

Consider a one-body state where the velocity field is $(\ref{p2572})$, i.e., $\rho_m\bu_{\pm} =
\pm \hbar\nabla\rho_m/2m$. This equation is Fick's law of diffusion \cite{Gillespie} with
diffusion coefficient ($D_\pm = \mp \hbar/2m$), but since (\ref{p0254}) is not in the proper form
of a continuity equation, the well known diffusion equation ($\partial\rho_m = D\nabla^2\rho_m$)
\cite{Gillespie} is not satisfied.


\alt{Moved material from Section~5.}

For the remainder of this section, the continuity equation~(\ref{p4288}) is shown to reduce the
the Poisson equation, and to the Laplace equation for stationary states, if the velocity fields are
orthogonal $\bv\cdot\bu = \mathbf{0}$. This material is not needed for the later results.

Next we examine the continuity equation~(\ref{p4288}), the one implied by the time-dependent
Sch\"odinger equation.  Let a sum of the subscript $i$ over $\{1,\cdots, n\}$ be understood, and
we suppress the $i$ subscript on the del operators $\nabla$.  The process of expanding out the
divergence term from (\ref{p4288}) and then using (\ref{p3312}) is
\[
\mathbf{0} = \partial \dens + \nabla\cdot(\dens\bv_i)  =
\partial \dens + \nabla\dens\cdot\nabla\bar{S} + \dens\nabla^2\bar{S},
\]
where the de Broglie velocity is $\nabla_i\bar{S} = \nabla_iS/m = \bv_i$.  Hence,
\begin{equation} \zlabel{4494}
  -\partial \dens = \nabla\dens\cdot\nabla\bar{S} + \dens\nabla^2\bar{S}, \quad \bar{S} \defi  S/m.
\end{equation}

\alt{P 17/L 28: Added an equation at the end, indicating that a condition given is equivalent
  to the two velocity fields being orthogonal. Removed mention of hydrogen atom states.}

We take this opportunity to obtain the continuity equation under two special conditions:
\begin{equation} \label{9904} 
  \text{If } \nabla\dens\cdot\nabla\bar{S} = \mathbf{0}, 
  \text{ then } \nabla^2\bar{S} = -\dens^{-1}\partial \dens = -\partial\ln\dens,
  \end{equation}
\begin{equation*} 
    \text{If } \nabla\dens\cdot\nabla\bar{S} = \mathbf{0} \text{ and } \partial\dens = 0, 
    \text{ then } \nabla^2 S = \mathbf{0}. 
\end{equation*}
These equations are the Poisson and Laplace equations, respectively.
A rule for taking the logarithm of a
dimensioned quantity, as in $\ln\dens$, is given in the second paragraph of the next section.
Note that the equation $\nabla\dens\cdot\nabla\bar{S} = \mathbf{0}$ is equivalent to the two
velocity fields being orthogonal, i.e., $\bv_i\cdot\bu_i = \mathbf{0}$, and this can be verified
by examining the definitions of the two velocity fields, Eq.~(\ref{p4720}) and (\ref{p3312}). 

\section{Momentum Potentials and the Kinetic Energy Fields\zlabel{2370}}

\alt{Defined ``momentae'' and the definition is used later.}

{\bf Definition.} The particle momentae set is ordered pair $\{m\bv,m\bu_\pm\}$, where the
momentae members are defined by Eqs.~(\ref{p3312}) and (\ref{p4720}).

\alt{P17/ L 38: Replaced the paragraph to improve clarity.}

In previous sections, definitions of the two particle-momentae, $m\bu_\pm$ and $m\bv$, are
made when these fields make an appearance in classical-mechanical equations, where the
classical-mechanical equations are implied by Schr\"odinger equations. In quantum mechanics,
momentum, an observable, is defined via axioms involving the momentum operator. Therefore, it
seems reasonable to investigate if rules can be found that yield the particle momentae using
the momentum operator, giving independent, and alternative, definitions of these fields that
do not involve Schr\"odinger equations.
This approach is investigated in this section, yielding two particle-moment definitions, as a
pair of fields, defined by a complex-valued function involving the momentum operator: One
particle momentum is taken as the real part of the function; one is taken as the
imaginary part. The resulting pair of momentums are the momentae, $m\bu$ and $m\bv$, defined
above. 
Furthermore, using the momentum operator, a natural definition for a kinetic energy-field is
also given, giving the same field as in Sec.~(\ref{p7835}).  To reduce clutter in the
derivation, we begin with a one-body state and use the down-hill velocity choice $\bu_-$.

Let $\zeta$ be a constant with the same units as the probability density $\rho$, i.e., per
volume.  Let the natural logarithmic function $\ln$ be refined by
$\ln\rho\,\dot{=}\,\log_e(\rho/\zeta)$, taking care of the requirement that the function $\ln$
is only defined on dimensionless quantities, and this definition can be used for other
dimensioned quantities with the modification understood. This $\ln$ definition is useful when
only derivatives of $\ln$ are assigned meaning. Let
\begin{equation} \zlabel{5977}
\theta_\pm = \pm\fc{\hbar}{2}\ln\rho. 
\end{equation}
Using this  definition, and the velocity definitions~(\ref{p3312}) and (\ref{p2572}), the
irrotational particle-momentae can be expressed using their potentials:
\begin{equation} \zlabel{5924}
m\bv = \nabla S,\qquad m\bu_\pm = \nabla\theta_\pm.
\end{equation}

\alt{P19/L 4: Added at the end of the paragraph that definition~(\ref{5200}) for velocity
  $\bu$ fixes its direction to within a sign.}

Let $\hat{\text{P}}$ be the momentum operator for quantum mechanical states of one-body
systems. Let the two particle-momentums of a state with wavefunction $\Psi$ be the real and
imaginary parts of $(\Psi^*\hat{\text{P}}\Psi)\rho^{-1}$, where $\rho = \Psi^*\Psi$ is the
probability distribution. Similarly, the fluid momentums per volume are defined as the real and
imaginary parts of $\Psi^*\hat{\text{P}}\Psi$.  With the definition
$i\hbar^{-1}\hat{\text{\text{P}}} \defi \nabla$ in mind, we obtain the following equation sequence:
\[
\Psi^*\nabla\Psi = i\hbar^{-1}\Psi^*\hat{\text{P}}\Psi =  R e^{-iS/\hbar}\nabla(R e^{iS/\hbar}) = i\hbar^{-1}\rho\nabla S + R\nabla R,
\]
\[i\hbar^{-1}\Psi^*\hat{\text{P}}\Psi = i\hbar^{-1}\rho\nabla S + \fc12\nabla\rho, \]
\[\Psi^*\hat{\text{\text{P}}}\Psi = \rho\nabla S - i\fc{\hbar}{2}\nabla\rho.\]  
Using the momentae potential definitions~(\ref{5924}), we obtain the objective:
\begin{equation} \zlabel{5200}
  \Psi^*\hat{\text{P}}\Psi = \rho\nabla S + i\rho\nabla\theta = \rho_m\bv + i\rho_m\bu,
\end{equation}
\begin{equation} \zlabel{5204}
   \fc{\Psi^*\hat{\text{P}}\Psi}{\Psi^*\Psi} = \nabla S + i\nabla\theta = m\bv + im\bu, 
\end{equation}
where $\rho_m \defi m\rho$, $\bu = \bu_-$, and we will continue using the downhill velocity
choice $\bu_-$.  Since any complex function can be written in polar form, the above formulae
for momentums $m\bu$ and $m\bv$ are determined by the momentum operator $\hat{\text{P}}$.
Since $\bu_-$ has the same sign as $\bv$ in the above equations, we use this as justification
for choosing $\bu_-$ over $\bu_+$. However, the corresponding $\bu_+$ equations are obtained
simply by replacing $\bu_-$ with $-\bu_+$. Note that the definition $\bv_\pm = \pm\nabla S/m$
corresponds to two linearly independent wavefunctions $Re^{\pm iS/\hbar}$, with the same $R$
function, if $S\neq \mathbf{0}$.
Also note that the particle-momentum definition~(\ref{5200}) for the Bernoullian velocity $\bu$
fixes the direction of this velocity field to within a sign, removing the flexibility in the
direction of $\bu$ for its corresponding kinetic energy, as pointed out in the end of
Sec.~\ref{p7835}. 

\alt{P 19 L 13: Sentence added at the end of the paragraph to make a point.}

Next we give a natural definition for the kinetic energy $K$, and obtain its formula:
\begin{equation}
  \zlabel{5208}
  \hspace{-8ex}
K \defi K_\bv + K_\bu \defi \fc{1}{2m}\left\vert\fc{\Psi^*\hat{\text{P}}\Psi}{\Psi^*\Psi}\right\vert^2 =
\fc12m v^2 + \fc12mu^2 = \fc{1}{2m} \vert\nabla S\vert^2 + \fc{1}{2m}\vert\nabla\theta_\pm\vert^2,
\end{equation}
and this equation also defines two kinetic energy formulae, $K_\bv$ and $K_\bu$. These
definitions yield the same fields as identified in previous sections, where both fields appear
in Eq.~(\ref{p5522}).

Since the generalization to the $n$-body case is trivial, and discussed below, the above
labeled three equations gives support for the interpretations for the velocities functions
present in the energy equation Eq.~(\ref{p5522}). It also suggests that the function $P_i$ from
Eq.~(\ref{p5522}) is not a kinetic energy term. The function $P_i$ in Eq.~(\ref{p5522}) is also
probably not one associated with a body force, since such potentials are usually universal,
and, therefore, would not have a formula that depends on the probability density, as in
definition~(\ref{p4722}).

The $n$-body case is easily obtained by the replacements $X \to X_i$ and $\nabla Y \to \nabla_i Y$ where
\[
X\in \{\hat{\text{P}}, \bv, \bu, K,K_\bv,K_\bu\},\quad \nabla Y\in \{\nabla S, \nabla\theta\},
\]
and where the equations still hold by summation over the dummy index $i$. For example,
Eq.~(\ref{5208}) becomes
\[ 
\sum_{i=1}^n K_i \defi \sum_{i=1}^n  K_{\bv i} + K_{\bu i} \defi
\fc{1}{2m}\sum_{i=1}^n  \left\vert\fc{\Psi^*\hat{\text{P}_i}\Psi}{\Psi^*\Psi}\right\vert^2
\hspace{15ex} 
\]
\begin{equation} \label{5210}
\hspace{5ex}  
      = \sum_{i=1}^n\left(\fc12m v_i^2 + \fc12mu_i^2\right)
      = \fc{1}{2m}\sum_{i=1}^n\left(\vert\nabla_i S\vert^2 + \fc{1}{2m}\vert\nabla_i\theta_i\vert^2\right).
\end{equation}
This equation also holds without the summations.

\alt{Moved ``A velocity compatibility problem to Sec.~(\ref{4382}).}

\section{Notation Involving Sets with the Same Cardinality \zlabel{5147}}

An examination of Eq.~(\ref{p4791}) indicates that there are three shortcomings in its
notation: 1) clutter caused by the summation symbols $\Sigma$, 2) eyestrain from the subscript
symbols $_i$ and $_j$, and 3) over accented from too many capital letters. In this section, we
introduce new notation to remove these shortcomings and put the $n$-body case on the same
footing as the one-body case, applicable almost all of the equations encountered in this work.

A product compatible collection $N\cdot$ is a family containing sets $A_1,A_2,\cdots$ with the same
cardinality. Also, the following operations are defined:
\[
\sum_{i\in I}x_i, \qquad \sum_{i\in I} x_i y_i, \quad x_i\in A\in N\cdot, \quad y_i\in B\in N\cdot,  
\]
where $I$ is an index set, and each of the members of $N\cdot$ are indexed by $I$. Since $x$
and $y$ are not defined elements of $A$, $B$, we define these symbols, \emph{in equations}, to be
\[
x \defi \sum_{i\in I}x_i, \qquad xy \defi \sum_{i\in I} x_i y_i.
\]
We also give another meaning to the symbols $x$ and $y$ when they are not in an equation: $x$
is the set $A$; $y$ is the set $B$, e.g., $x =\{x_i \vert i \in I\}$. Hence, it is not
necessary to use the symbols $A$ and $B$. Also, if the cardinality is one, then the single
element $x_1$ from the set $x$ is also denoted by $x$. This convention is also used for $f(C)$,
the image of the set $C$ under the function $f$ by writing $f(x)$, if $C = \{x\}$, instead of
the notation $f(\{x\})$.  It can also be convenient to be able to append $N\cdot$ with
elements. So we define $N\cdot\defi N\cdot + z$ to mean that $N\cdot\cup\{z\}$ is now the new
definition of $N\cdot$, and this can be understood to have been done, when there is no
misunderstanding.

\alt{P 21 L 21 Added an additional example involving $\vert\nabla S\vert^2$ for clarity, displayed below.}

To apply this notation to Eq.~(\ref{5210}), we define a member of $N\cdot$ for each variable
with an $i$ subscript, and these variables are indexed by ($I = \{1,\cdots n\}$).  Examples
being ($K = \{K_i \vert i \in I\}$), ($u^2 = \{u_i^2 \vert i \in I\}$), ($\nabla S =
\{\nabla_iS \vert i \in I\}$), and we also use
\[
\vert\nabla S\vert^2 = \nabla S\cdot\nabla S \defi \sum_i\nabla S_i\cdot \nabla S_i.
\]
The result of this notation change is that Eqs.~(\ref{5210}) and (\ref{5208})
represent the same equation. Also the application of notation to (\ref{p4791b}) gives
(\ref{p2574}), and (\ref{p5522}) becomes
\begin{equation} \zlabel{p5544}
-\partial S = \fc{1}{2}m v^2 + \fc12 m u^2 + P\rho^{-1} + U,
\end{equation}
where $U$ does not use this notation, being still defined by Eq.~(\ref{p0025}).
In this equation, since the left-hand side cannot be expressed as a sum over $n$ terms, only
the right-hand side is summed over the index set $I$, except for $U$. In order that a summation
can be given explicitly, if it is useful to do so, let the operator $\Sigma$ be defined by
\[
\sum x \defi \sum_{i\in I}x_i, \qquad \sum xy \defi \sum_{i\in I} x_i y_i, \quad \text{and so on.}
\]
giving 
\[
-\partial S = U + \sum\hspace{0.2ex} \fc{1}{2}m v^2 + \fc12 m u^2 + P\rho^{-1},
\]
where $\sum$ acts on all terms on its right side, until either an equal sign or the equation end.

In the above equation and Eq.~(\ref{p5544}), the over accented $\Upsilon$ is now $\rho$, and
$P$ could be changed to $p$.  I find that subscripts get in the way of the process of the
evaluation of an equation. For subscripted equations, in the process of weeding out the
clutter, I find myself staring at the little subscripts, instead of taking in the whole
equation all at once.  I also find that two many capital symbols, especially ones in Greek,
make it difficult to distinguish and classify the factors of terms in equations. This may be
the reason that functional analysis monographs usually use symbols like $x$ and $y$ for maps
and reserve capitals for sets.

\section{Energies \zlabel{2371}}

\alt{P 22 L 14 Added introductory paragraph and modified wording for the paragraph that follows,
including a definition.}

In this section, as in the particle momentae definitions in Sec.~\ref{2370}, two energy fields
are defined as the real and imaginary parts of the function $\Psi^*\hat{H}\Psi/\rho$, where 
$\hat{H}$ is the Hamiltonian (energy) operator. Using the Schr\"odinger Eq.~($i\Psi^*\partial\Psi =
\Psi^*\hat{H}\Psi$), the approach yields a pair of energy fields as the real and imaginary
parts of $i\Psi^*\partial\Psi/\rho$: ($\bE _S = -\partial S$) and ($\bE _\theta =
-\partial\theta$), where $S$ and $\theta$ are the momentae potentials, defined by
Eq.~(\ref{5924}).
The energy field $(\bE_S= -\partial S)$ is substituted into the Hamilton-Jacobi
Eq.~(\ref{p5544}) of Sec.~\ref{p4242}, giving an energy equation, Eq.~(\ref{5830c}). In the
special case of time independence, the energy field $E_S$ reduces to the energy eigenvalue $E$
of the Schr\"odinger equation. Later in Sec.~(\ref{4379}), the other energy field $\bE_\theta$,
an unknown, makes an appearance in an energy form of the continuity equation, an equation
equivalent to the Bohmian-mechanics continuity Eq.~(\ref{p4288}).

Let the energy functions $\bE _S$ and $\bE _\theta$ be the real and imaginary parts of the
function $ \rho^{-1}\Psi\hat{H}\Psi$, where the Schr\"odinger
Eq.~(\ref{p2588}), can be written
\begin{equation} \zlabel{5882}
    i\hbar\rho^{-1}\Psi^*\partial\Psi = \rho^{-1}\Psi\hat{H}\Psi.
\end{equation}
Setting $\Psi = \phi e^{iS/\hbar}$, and working on the left-hand side of (\ref{5882}), we
obtain the following equation sequence:
\[ e^{-iS/\hbar}\phi \partial(\phi e^{iS/\hbar}) = \phi\partial\phi + i\rho\hbar^{-1}\partial S, \]
\[  e^{-iS/\hbar}\phi \partial(\phi e^{iS/\hbar}) = \fc12\partial\rho + i\rho\hbar^{-1}\partial S,\]
\[ i\hbar e^{-iS/\hbar}\phi \partial(\phi e^{iS/\hbar}) = -\rho\partial S  +i\fc{\hbar}{2}\partial\rho,\]
\begin{equation} \zlabel{4383}
  i\hbar\rho^{-1}\Psi^*\partial\Psi = -\partial S + i\fc{\hbar}{2}\partial\ln\rho.
\end{equation}
Using definition~(\ref{5977}) with $\theta = \theta_-$ and Eq.~(\ref{5882}), we discover that 
\begin{equation}
\zlabel{7204a} \rho^{-1}\Psi\hat{H}\Psi = i\hbar\rho^{-1}\Psi^*\partial\Psi
     = -\partial S - i\partial\theta,
\end{equation}
and the above definitions give the desired result:
\begin{equation} \zlabel{5830}
  \bE _S = -\partial S, \quad  \bE _\theta = -\partial\theta = \fc{\hbar}{2}\partial\ln\rho,
\end{equation}
such that
\begin{equation} \zlabel{5828}
  \fc{i\hbar\Psi^*\partial\Psi}{\Psi^*\Psi} = \fc{\Psi^*\hat{H}\Psi}{\Psi^*\Psi} = \bE _S + i\bE _\theta,
\end{equation}
and $\partial\ln\rho = \rho^{-1}\partial\rho$ is used in Eq.~(\ref{4383}).  It is worth noting
that, since $\partial\ln\rho = \rho^{-1}\partial\rho$, the right-hand side equation of
(\ref{5830}), can be written
\begin{equation} \zlabel{5830b}
\bE _\theta\rho = -\rho\partial\theta = \fc{\hbar}{2}\partial \rho.
\end{equation}
Equation~(\ref{5830}) gives a further support for the choice $\theta = \theta_-$, since the
corresponding function, $-\partial S$, has a minus sign. Henceforth, in part to reduce clutter,
we use the choice $\theta = \theta_-$, giving the downhill choice $\bu_-$.  Also, substituting
(\ref{5830}) into (\ref{p5544}) with the $P$ notation changed to $P_\bu$, we have
\begin{equation} \zlabel{5830c}
\bE _S = \fc{1}{2}m v^2 + \fc12 m u^2 + P_\bu\rho^{-1} + U,
\end{equation}
and here we use the notation from Sec.~\ref{5147} for $n$-bodies.  Since Eq.~(\ref{p5544}),
together with the equation~(\ref{p4288}), are equivalent to the Schr\"odinger' (\ref{p2588}),
the above equation defines an energy field $\bE _S$, not necessarily uniform and constant in
time, that is assigned to time-dependent systems that satisfy the time-dependent Sch\"odinger
equation~(\ref{p2588}).

\alt{P23 /L 23: Added a two sentence paragraph to assist reader in using the notation.}

It is worth noting that, to return an equation like Eq.~(\ref{5830c}) to the subscripted
notation, a subscript, say $j$, is appended to the pressure and velocity (or speed)
fields---$u^2$, $v^2$, and $P_\bu$ in Eq.~(\ref{5830c})---followed by a summation over the dummy
index $j$. Using this rule makes the derivations of Sec.~(\ref{p1830}) easier to follow.



\alt{P 23/L 25: The next section is split into two sections and introductory paragraphs are added to
  the two sections.}

\section{Pressures \zlabel{2475}}


\alt{P 23 L 27. Split the paragraph into two, added introductory material, and the modified wording
  for the second paragraph.}

This section extends the method of definition for energies and momentae fields to pressures.
This physical property has not been defined as an observable for quantum mechanics, and,
therefore, it does not have a linear-operator assignment.  Definition~(\ref{4007}), involving the
divergence of momentum density, yields the same pressure $P$ as in the previous sections,
defined in Eqs.~(\ref{p2572}) or (\ref{p4722}), where the notation is changed from $P$ to
$P_\bu$. The second pressure $P_\bv$, an unknown, appears in an energy form of the
continuity equation~(\ref{5037n}) from Sec.~\ref{4379}, along with the energy field $E_\theta$.

We begin with the notation from Sec.~\ref{5147} and then switch to the subscripted notation in the
next section.  We also continue to use the choice $\bu = \bu_-$.
Let the the pressures, $P_\bu$ and $P_\bv$, be defined, in the following manner, where they are
proportional to the divergence of a momentum density:
\begin{equation} \zlabel{4002}
P_\bu \defi \fc{\hbar}{2m}\nabla\cdot\rho_m\bu_-,\qquad  P_\bv \defi \fc{\hbar}{2m}\nabla\cdot\rho_m(-\bv).
\end{equation}
Using definition~(\ref{p2572}) for the velocity $\bu_-$, we have $\rho_m\bu_- =
-(\hbar/2)\nabla\rho$, giving the previous definition~(\ref{p2572}) for the pressure with
$P_\bu = P$:
\begin{equation} \zlabel{4005}
P_\bu \defi \fc{\hbar}{2m}\nabla\cdot\rho_m\bu_- = \fc{\hbar}{2m}\nabla\cdot\left(-\fc{\hbar}{2}\nabla\rho\right)
= -\fc{\hbar^2}{4m}\nabla^2\rho.
\end{equation}

Using definitions (\ref{4002}) and (\ref{5200}), $\Psi^*\hat{\text{P}}\Psi = \rho_m\bv +
i\rho_m\bu$, we obtain an alternate definition for the pressures:
\begin{equation} \zlabel{4007}
\fc{\hbar}{2m}\nabla\cdot\left(\Psi^*\hat{\text{P}}\Psi\right) = -P_\bv + i P_\bu.
\end{equation}
The sign difference between the two pressures suggests that one of these pressures would be
better with a sign change, and the resulting function could be considered a tension. However,
since the pressures, especially $P_\bu$, are not thermodynamic pressures, i.e., they can be negative valued, such a
sign change in the equations seems to make little or no difference in the given
interpretations.
The other possibility is to use the uphill velocity choice $\bu_+$.

 \section{Equalities involving Time Derivatives and Pressures \zlabel{4379}}

This section uses the continuity equation to derive a number of relations involving the
pressures, most having time derivatives. An energy equation~(\ref{5037n}) is obtained,
containing both the unknown pressure $P_\bv$ and energy $\bE_\theta$ fields, and this equation
is shown to be equivalent to the continuity equation~(\ref{p4288}). At the conclusion of this
section, we have a total of three defined ordered pairs---$\{m\bv,m\bu\}$, $\{P_\bv,P_\bu\}$,
and $\{\bE_S,\bE_\theta\}$---that are present in two main equations, (\ref{5830c}) and
(\ref{5037n}). These two equations, along with the definitions of the three ordered pairs of
fields, are equivalent to the time-dependent Schr\"odinger equation~(\ref{p2588}), indicating
that the method is ab initio.  Each and every term is identified from the two equations, one
term $U$ being the classical electrostatic terms~(\ref{p0025}), indicating the the approach is
complete in its energy description.

\alt{P 24/L 23: Switched the order in two equations: (\ref{4302n}) and the above (\ref{4302n}).}

\alt{P 24/L 32: Mentioned that (\ref{4302n}) is a form of the continuity equation.}


\alt{P 24/L 42: Mention that (\ref{5025n}) is used in the next section.}

\alt{P 25/ L 7: Fixed an error in the interpretation of (\ref{5025n}) at the end.}

In what follows in this section, because the continuity equation is used, it is necessary to use
the subscripted notation. For example, the above definitions are
\begin{equation} \zlabel{4005n}
  P_{\bu i} \defi \fc{\hbar}{2m}\nabla_i\cdot\rho_m\bu_{i-},\qquad
  P_{\bv i} \defi \fc{\hbar}{2m}\nabla_i\cdot\rho_m(-\bv_i).
\end{equation}
For the pressure $P_{\bv i}$, we multiply the continuity Eq.~(\ref{p4288}), using the notation
$\rho = \Upsilon$, by $\hbar/2$ and $m/m$, giving
\[
\fc{\hbar}{2}\partial\rho = -\fc{\hbar}{2m}\sum_i^n\nabla_i\cdot(\rho_m\bv_i),
\]
and it follows from the above definition~(\ref{4005n}) for $P_{\bv i}$ that
\begin{equation} \zlabel{4302n}
  \fc{\hbar}{2}\partial\rho = \sum_i^n P_{\bv i},
\end{equation}
a form of the continuity equation.  Taking a gradient of this result and switching the order, we have
\[
\sum_i \nabla_j P_{\bv i} = \fc{\hbar}{2}\partial(\nabla_j\rho).
\]
Using the equality $\nabla\rho_j = -(2/\hbar)\bu_{j-}\rho_m$ from Eq.~(\ref{p2572}), and
interchanging the dummy indices, we obtain the additional result:
%
%
\begin{equation} \zlabel{5025n}
\sum_j \nabla_i P_{\bv j} =-\partial(\rho_m\bu_i), \quad i \in\{1,\cdots n\}.
\end{equation}
This equation is used in the next section.  For the special case of one body, the resulting
equation,
\[
-\nabla P_{\bv} = \partial(\rho_m\bu), 
\]
is \emph{not} Newton's second law, since, as discussed in the next section after
Eq.~(\ref{4887}), the partial time-derivative is a local derivative, assigned to fixed
points in the field, instead of a time derivative from an equation of motion that is assigned
to a particle, as it moves through space.

\alt{P 24/L 38: Added the next single-line equation to make a derivation easier to follow.}

\alt{P 25/L 14: Mentioned that (\ref{4291}) is the Poisson equation for the one-body case.}

Applying the operator $(\hbar/2m)\nabla_i\cdot$ to Eq.~(\ref{5025n}),
\[  \fc{\hbar}{2m}\sum_j \nabla_i^2 P_{\bv j} = -\fc{\hbar}{2m}\partial[\nabla_i\cdot(\rho_m\bu_i)],\]
and using the definition~(\ref{4005n}), $P_{\bu i} = (\hbar/2m)\nabla_i\cdot\rho_m\bu_i$, and
switching the order, we discover that the two types of pressures are related by
\begin{equation} \zlabel{4291}
  \partial P_{\bu i} =  -\fc{\hbar}{2m} \sum_j \nabla_i^2 P_{\bv j},
\end{equation}
the Poisson equation for the one-body case.

\alt{P 26/L 17: Added an inline equation to make the assist the reader.}

For an energy equation for $\bE _\theta$, we substitute (\ref{5830}), written
($(\,\hbar/2)\partial\rho = \bE _\theta\rho$), into~(\ref{4302n}), ($(\,\hbar/ 2)\partial\rho = P_{\bv
  1} + \cdots P_{\bv n}$), giving
\begin{equation} \zlabel{5037n}
  \bE _\theta = \rho^{-1} \sum_{i}^n P_{\bv i}.
\end{equation}
Hence, $\bE _\theta\rho$ is a pressure with corresponding compression energy $\bE_\theta$.

\alt{P 25/L 20: Paragraph is added that demonstrates that Eq.~(52) is equivalent to the continuity Eq.~(20).}

Since the derivation of (\ref{5037n}) from (\ref{p4288}) is reversible, Eq.~(\ref{5037n}) is
mathematically equivalent to the continuity equation (\ref{p4288}). The reversibility is proven
by $\mathbf{1)}$ doing the derivation in reverse, where $P_{\bv i}$ is nothing more then a
mathematical field, defined by the right-hand side of Eq.~(\ref{4005n}), and the field $\bv_i$,
present in that equation, is defined by (\ref{p4720}). By $\mathbf{2)}$ noting that the
satisfaction of the field $E_\theta$, in Eq.~(\ref{5830}), follows from its mathematical
definition~(\ref{5830}).

\alt{The next sentence is added:}

Equation~(\ref{5037n}) is used in the next section.

\section{The Total Energy and Euler Equation \zlabel{p1830}}

\alt{P 24/L 24. Added introductory paragraph.}

This section combines the two energy equations, Eqs.~(\ref{5830c}) and (\ref{5037n}), giving a
total-energy equation~(\ref{5830d}), containing all three ordered pairs of fields.  An $n$-body
generalization of the Euler equation of fluid dynamics is derived from this equation,
Eq.~(\ref{8888d}), and this derivation is much longer than previous ones. This Eulerian
equation is viewed as a sum of two equations corresponding to two interacting systems. The
\emph{local} time derivative of the velocity fields are shown to satisfy Eqs.~(\ref{4887}),
where, for the case of spin-free one-body states, $\partial\bv(\mrr,t)$ and $\partial\bu(\mrr,t)$
are the time rate of change of the velocity fields at time $t$ and \emph{fixed} point
$\mrr\in\mathbb{R}$.

%
%

Combining Eq.~(\ref{5830c}) and (\ref{5037n}), for the one-body case, we obtain a definition
of the total energy $\bE$:
\begin{equation} \zlabel{5830d}
  \bE  \defi \bE _S + \bE _\theta =  \fc{1}{2}m v^2 + P_\bv\rho^{-1} + \fc12 m u^2 + P_\bu\rho^{-1} + U,
\end{equation}
and an alternative worth mention is $\bE  \defi \vert\bE _S + i\bE _\theta\vert$.  The above
equation holds for $n$-bodies using the notation from Sec.~\ref{5147}. Using the
subscripted notation, it can also be derived using equations (\ref{5830}), (\ref{p5522}), and
(\ref{5037n}).

\alt{P 25/L 41: Added one word to correct grammar.}

\alt{P 26/L 7: Mentioned equalities from Sec,~(\ref{p4242}) involving continuity equation.}

Returning to Eq.~(\ref{5830d}), in the special case where $\bv\cdot\bu = \mathbf{0}$, we can write
\begin{equation} \zlabel{5830g}
  \fc{1}{2}m w^2 + P\rho^{-1} + U = \bE ,
\end{equation}
where
\[
w^2 = \mathbf{w}\cdot\bw, \quad \bw = \bu + \bv, \qquad P = P_\bv + P_\bu. 
\]
For a one-body system, Eq.~(\ref{5830d}) has the same form as the Bernoullian'
(\ref{p2574}). While one objective of this work is to avoid complex-valued physical-properties,
it is still worth mentioning that the definition
\[
w^2 = \vert\bv + i\bu\vert^2,
\]
gives (\ref{5830g}) from (\ref{5830d}) for the general case, where $\bv\cdot\bu \ne \mathbf{0}$
is permitted.  The consequences of $\bv\cdot\bu = \mathbf{0}$ for the continuity equation is
given at the end of section~\ref{p4242}.

\alt{P 26/L 7: A new paragraph is added describing the meaning of Eq.~(\ref{4887}), and this equation and
its derivation is moved up into this paragraph.}

Next we derived a pair of equations involving $\partial\bv$ and $\partial\bu$, needed for the
derivation of an Eulerian equation that follows. This is accomplished by taking the gradient of
$\bE _S = -\partial S$ and $\bE _\theta = -\partial\theta$, i.e., Eq.~(\ref{5830}), and then
using the particle momentum definitions~(\ref{5924}), giving
\begin{equation} \zlabel{4887}
  -\nabla\bE _S = m\partial\bv,\quad -\nabla\bE _\theta = m\partial\bu.
\end{equation}
Despite the forms of these equations, they are \emph{not} to be confused with Newton's second
law for particles with conservative forces $-\nabla\bE _S$ and $-\nabla\bE _\theta$,
respectively.
In classical fluid dynamics \cite{Munson}, $\partial\bv$ and $\partial\bu$ are \emph{local} accelerations:
For a fixed position in space $\mrr\in\mathbb{R}$ and velocity field $\bu$, $\partial\bu(\mrr,t)$
is the time rate of change of the velocity fields at point $\mrr$ and time $t$. For steady flow,
these partial derivatives vanish, and the acceleration of a fluid particle is obtained by the
time derivative of the composite, $\bu = \bu(\mrr)$, such that $\mrr = \mrr(t)$. For the case of
nonsteady flow, we have $\bu = \bu(\mrr,t)$, with $\mrr = \mrr(t)$, where the nonpartial, total
derivative of $\bu$ is called the material derivative \cite{Munson,Shapiro,Kelly}.

\alt{P 26/L 34: removed not needed parts of the second equations in the sequence.}

\alt{P 26/L 41 and P 28 L 19: mentioned that Eq.~(\ref{8888c}) represents $n$ such equations.}

Next we derive an Euler equation for one-body systems, and, separately, obtain the $n$-body
equation. We begin by taking the gradient of (\ref{5830c}),
\begin{equation} \zlabel{4884}
\nabla\bE _S = \fc{1}{2}m\nabla v^2 + \fc12 m \nabla u^2 + \nabla\left(P_\bu\rho^{-1}\right) + \nabla U. 
\end{equation}
Substituting the left equation of (\ref{4887}) into Eq.~(\ref{4884}), and then adding the resulting equation to
Eq.~(\ref{5025n}), in the form $\rho^{-1}\partial(\rho_m\bu) + \rho^{-1}\nabla P_\bv = \mathbf{0}$,
we obtain the desired result for one-body systems:
\begin{equation} \label{8888}
    \hspace{-5ex}
  m\partial\bv  + \rho^{-1}\partial(\rho_m\bu) + \fc{1}{2}m\nabla\left(v^2 + u^2\right)
  + \rho^{-1}\nabla P_\bv + \nabla\left(P_\bu\rho^{-1}\right) + \nabla U  = \mathbf{0}.
\end{equation}
For the $n$-body case, we begin with Eq.~(\ref{5830c}), return to the subscripted notation
with summation index $j$, take a gradient $\nabla_i$ of that equation, and then follow the
procedure above for the one-body case with $\nabla$ replaced by $\nabla_i$, giving the
following equation sequence:
\[\nabla_i\bE _S = \fc{1}{2}m \nabla_iv_j^2 + \fc12 m \nabla_iu_j^2 + \nabla_i\left(P_{\bu j} \rho^{-1}\right) + \nabla_iU,\]
\[ -\nabla_i\bE _S = m\partial\bv_i,\quad -\nabla_i\bE _\theta = m\partial\bu_i,\]
\begin{equation*} \lab{5025n}
  \rho^{-1}\partial(\rho_m\bu_i) + \rho^{-1}\mbox{$\sum_j$} \nabla_i P_{\bv j} = \mathbf{0},
\end{equation*}
\begin{equation}
  \label{8888c}
  \hspace{-9ex}
  m\partial\bv_i + \rho^{-1}\partial(\rho_m\bu_i) + \fc{1}{2}m \nabla_i\left(v_j^2 + u_j^2\right)
  + \rho^{-1}\nabla_i P_{\bv j} + \nabla_i\left(P_{\bu j}\rho^{-1}\right) = -\nabla_iU,
\end{equation}
where there is an understood sum over index $j$ and $i\in\{1,\cdots n\}$.

\alt{P 26/L 43: Copied the first mentioned equation from the appendix as an aid to the reader
  and defined the terms.}

Next, we put the one-body equation (\ref{8888}) in a form so that it can be
compared to the variable mass Euler equation~(\ref{7288}),
\[
\pa{}{t}(\rho_m\bu) + \fc12\rho_m\nabla u^2 + \nabla\cdot(\rho_m\bu)\bu + \nabla p + q\rho\nabla\Phi = 0,
\]
derived in~\ref{p8250}, with pressure $p$, force per charge $(-\nabla\Phi)$,
mass density $\rho_m$, and charge density $q\rho$.  First note the following identity used
below:
\begin{equation} \zlabel{2748}
  \rho\nabla\left(\fc{P}{\rho}\right) = -P\fc{\nabla\rho}{\rho} + \nabla P. 
\end{equation}
Multiplying Eqs.~(\ref{4002}) and (\ref{p2572}), given by
\[
\nabla\cdot\rho_m\bu = \fc{2m}{\hbar}P_\bu, 
\qquad \bu = -\fc{\hbar}{2m}\fc{\nabla\rho}{\rho}, \;\;\; \bu = \bu_-,
\]
we obtain
\[
\nabla\cdot(\rho_m\bu)\bu = -P_\bu\fc{\nabla\rho}{\rho}.
\]
Substituting this one into identity (\ref{2748}) gives the desired equality for $P_\bu$:
\begin{equation} \zlabel{2077}
  \rho\nabla\left(\fc{P_\bu}{\rho}\right) = \nabla\cdot(\rho_m\bu)\bu + \nabla P_\bu.
\end{equation}
Substituting this equation into the Eq.~(\ref{8888}), after multiplying the equation by $\rho$, gives the final result: 
%
%
\begin{equation} \zlabel{8888b}
  \hspace{-9ex}
 \rho_m\partial\bv  + \partial(\rho_m\bu) + \fc{1}{2}\rho_m\nabla \left(v^2 + u^2\right)
 + \nabla \left(P_\bv + P_\bu\right) + \nabla\cdot(\rho_m\bu)\bu   + \rho\nabla U  = \mathbf{0}.
\end{equation}

\alt{P 27/L 30 Removed ``= 0'' error for equation sum and put it in in-line.}

\alt{P 27 /L 35 The two Euler equations are given in a more specialized form, with a mention of the general one.}

In order to assign a meaning to portions of Eq.~(\ref{8888b}), we write the equation as a sum of two equations:
($\mathbb{EQ}_\bu + \mathbb{EQ}_\bv$),
where
\[\mathbb{EQ}_\bu \equiv \partial(\rho_m\bu) + \fc12 \rho_m\nabla u^2 + \nabla\cdot(\rho_m\bu)\bu + \nabla P_\bu  + \rho\nabla U  = \mathbf{0},\]
\[\mathbb{EQ}_\bv \equiv \rho_m\partial\bv \; +\; \fc{1}{2}\rho_m\nabla v^2  + \nabla P_\bv = \mathbf{0}.\]
For one-body systems, these are Euler equations of fluid dynamics: Equation $\mathbb{EQ}_\bu$ is
the Euler Equation~(\ref{7288}) with variable mass; Equation $\mathbb{EQ}_\bu$ is the Euler
Equation~(\ref{7288b}) with conserved mass, and there is flexibility in the placement of the
body forces (per volume) $\rho\nabla U$. Both equations are for irrotational flow.

\alt{A condensed form of an equation is given, Eq.~(\ref{8888d}).  Below that equation, a
  sentences is added about the symmetry of the equation.}

For the $n$-body case, the derivation of the generalization of Eq.~(\ref{2077}) is
\[ \rho\nabla_i\left(\fc{P_j}{\rho}\right) = -P_j\fc{\nabla_i\rho}{\rho} + \nabla_i P_j,\]
\[
\lab{4005n} \nabla_j\cdot\rho_m\bu_j = \fc{2m}{\hbar}P_{\bu j}, 
\qquad \bu_i = -\fc{\hbar}{2m}\fc{\nabla_i\rho}{\rho}, \;\;\; \bu = \bu_-,
\]
\[\nabla_j\cdot(\rho_m\bu_j)\bu_i = -P_{\bu j}\fc{\nabla_i\rho}{\rho}, \]
\[\rho\nabla_i\left(\fc{P_{\bu j}}{\rho}\right) = \nabla\cdot(\rho_m\bu_j)\bu_i + \nabla_i P_{\bu j},\]
giving from (\ref{8888c}),
%
%
\begin{equation*}
  \hspace{-9ex}
 \rho_m\partial\bv_i  + \partial(\rho_m\bu_i) + \fc{1}{2}\rho_m\nabla_i \left(v_j^2 + u_j^2\right)
 + \nabla_i\left(P_{\bv j} + P_{\bu j}\right)  + \nabla_j\cdot(\rho_m\bu_j)\bu_i  = - \rho\nabla_i U,  
\end{equation*}
where there is an understood summation over index $j$ and $i\in\{1,\cdots n\}$.
Note that we can still use the notation system for Sec~\ref{5147} for the $j$ dummy index:
\begin{equation} \label{8888d} 
  \hspace{-2ex}
   \rho_m\partial\bv_{\mbox{\tiny $\!\!\triangle$}}  + \partial(\rho_m\bu_{\mbox{\tiny $\triangle$}}) + \fc{1}{2}\rho_m\nabla\vert w\vert^2
 + \nabla_{\mbox{\tiny $\!\!\triangle$}}\! P  + \nabla\cdot(\rho_m\bu)\bu_{\mbox{\tiny $\triangle$}}  = -\rho\nabla_{\mbox{\tiny $\!\!\triangle$}}\! U,
\end{equation}
where $\triangle \in\{1,\cdots n\}$, $w = v + iu$ and $P = P_{\bv} + P_{\bu}$. This equation lacks symmetry, with
respect to the interchange of $\bv$ and $\bu$, because one of these velocities satisfies a
continuity equation and one does not.  The symmetry is recovered by adding the continuity
equation $(\partial\rho + \nabla\cdot\rho\bv)\bv_i = \mathbf{0}$ satisfied by $\bv_i$.


\section{A Linear Combination of Eigenfunctions\zlabel{4290}}

\alt{P 28/L 26 Minor grammatical corrections and clarifications in the paragraph. Pointed out that
  the external potential is time independent.}

In this section, we investigate the energy fields $\bE_s$ and $\bE_\theta$, given by
Eqs.~(\ref{5830}) and (\ref{5830c}), for energy conservation.  Wavefunctions are considered
that are linear combinations of basis functions, where each function from the linear
combination is, separately, an eigenfunction of the same Hamiltonian operator with a
time-independent external potential. If members of such a linear combination are not degenerate,
the linear combination still satisfies the time dependent Schr\"odinger equation (\ref{p2588}),
but the linear combination does \emph{not} satisfy the time-independent Schr\"odinger
equation. In other words, the linear combination is not an eigenfunction of the Hamiltonian
operator.  We show that these quantities have the important property of being conserved over
all space for each moment of time. In other words, while energy flows from one region of space
to another, the total energy, over all space, is fixed in time.

\alt{P 28/L 41: Added $i\in\{1,2\}$ and removed ``and''.}

Using atomic units, let $\phi_1$ and $\phi_2$ be real-valued orthonormal functions that are
eigenfunctions of a Hamiltonian operator $\hat{H}$. Let
\[ \hspace{-2ex}
\psi_i(t) = e^{-i\veps_i t},\;\;
\vert C_1\vert^2 + \vert C_2\vert^2 = 1,\; C_i \in\mathbb{R},\; \rho_i\defi \phi_i\phi_i,\; i\in\{1,2\},\; \beta\,\dot{=}\,C_1C_2.
\]
In the following equation sequence, we start with the linear combination wavefunction $\Psi$,
and calculate the left-hand side of Eq.~(\ref{5828}), multiplied by $\rho =\sPsi\sPsi^*$, for a $n$-body
system with spin variables suppressed:
\[\sPsi(\mrr,t) \defi C_1\phi_1(\mrr)\psi_1(t) + C_2\phi_2(\mrr)\psi_2(t),\]
\[ i\partial\sPsi = C_1\veps_1\phi_1\psi_1 + C_2\veps_2\phi_2\psi_2,\]
\[\sPsi^* i\partial\sPsi =(C_1(\phi_1\psi_1)^* + C_2(\phi_2\psi_2)^*)(C_1\veps_1\phi_1\psi_1 + C_2\veps_2\phi_2\psi_2),\]
\[\sPsi^* i\partial\sPsi = C_1^2\veps_1\rho_1 + C_2^2\veps_2\rho_2 + \beta(\veps_1\phi_1\phi_2^*\psi_2^*\psi_1 + \veps_2\phi_2\phi_1^*\psi_1^*\psi_2),\]
\[
  \sPsi^*i\partial\sPsi = C_1^2\veps_1\rho_1 + C_2^2\veps_2\rho_2
  + \beta\phi_1\phi_2\left(\veps_1e^{i(\veps_2-\veps_1)t} + \veps_2 e^{-i(\veps_2-\veps_1)t}\right).
\] 
Using (\ref{5828}), we obtain the desired energy fields per volume:
\begin{equation}\zlabel{2203} 
  \bE _\theta\rho = \beta\phi_1\phi_2(\veps_2-\veps_1)\sin[(\veps_1-\veps_2)t],
\end{equation}
\begin{equation}\zlabel{2204} 
   \bE _S\rho = C_1^2\veps_1\rho_1 + C_2^2\veps_2\rho_2 + \beta\phi_1\phi_2(\veps_1+\veps_2)\cos[(\veps_2-\veps_1)t].
\end{equation}
Since the functions $\phi_1$ and $\phi_2$ are orthonormal, the spacial energy averages, 
obtained by integrating (\ref{2203}) and (\ref{2204}) over all space $\mathbb{R}^{3n}$ and
summing over all spin variables, are
\[
\bE _\theta(\text{avg}) = \mathbf{0}, \quad \text{and}\quad \bE _S(\text{avg}) =
C_1^2\veps_1 + C_2^2\veps_2.\]
Since these are constants, independent of time, both of these energies are conserved over all
space.

\alt{P29/L 35: Removed, not needed, last sentence in the paragraph.} 

If we repeat the derivation above with the following change $C_2 \to -C_2$, then the sign of
the $\beta$ terms in Eq.~(\ref{2203}) and (\ref{2204}) change. If, instead, we make the
following change $C_2 \to iC_2$, nothing dramatic happens: The $\beta$ terms in
Eq.~(\ref{2204}), for $\bE _S$, and Eq.~(\ref{2203}), for $\bE _\theta$, switch places.

For the general case, using a set of $m$ orthonormal spatial wavefunctions
$\{\phi_1,\cdots \phi_m\}$ that are eigenfunctions of a single Hamiltonian operator, the same
result is obtained: $\bE _S$ and $\bE _\theta$ are conserved over space. This result is
easily seen to follow because, with no loss of generality, the members of the set
$\{\phi_1,\cdots \phi_m\}$ are mutually orthogonal. The same result is obtained for a
wavefunction defined by an infinite sequence of orthonormal functions, since each term of the
corresponding sequences, for both $\bE _S$ and $\bE _\theta$, are conserved.


\alt{Reorganized this section and combined Sec.~(\ref{2370}) with (12).}

\section{Discussion of issues of velocity \zlabel{5382}} 

\alt{Introduction paragraph added.}

This section considers issues involving velocity. Sec.~\ref{4382} investigates a velocity
compatibility problem for the two velocity fields as a vector sum $\bv + \bu$.  Sec.~\ref{1025}
considers other possibilities for the Bernoullian velocity $\bu$.  Various issues of local
variable mass, angular momentum, spin, and kinetic-energy satisfaction are considered, for both
fluid and particle descriptions. The $n$s states of hydrogen atom are considered for the
investigation. The objective is to determine what velocity is the best match for the
Bernoullian velocity $\bu$, and to summarize the information available. For the most part, only
one-body systems are considered.

{\alt{Moved material from the end of Sec.~(\ref{2370}) to the next subsection and rewrote much
    of it with additions.}

\subsection{A velocity compatibility problem  \zlabel{4382}}


For the case of one-body, with a time-independent complex-valued wavefunction, let the velocity
be $\dot{\mathbf{q}} = \bu + \bv$. Since
\[
\fc12 m \dot{\mathbf{q}}\cdot\dot{\mathbf{q}} \ne \fc12m\bu\cdot\bu + \fc12m\bv\cdot\bv, 
\]
if $\bu\cdot\bv \ne 0$, the single particle does not have the correct kinetic energy.  Hence,
the two kinetic energies seem to represent different forms of energy.  Another justification,
for this posit, is that the Eulerian Eq.~(\ref{8888b}) is consistent with two interacting
``substates,'' where each one can have their own energy forms.  One possibility for the
kinetic-energy identification, involving the Bernoullian velocity, is that the vector
$\vert\bu\vert$ represents an average speed of Brownian-like motion, which, given the direction
of the velocity vector $\bu$ from Eq.~(\ref{p2572}), might be restricted to one-dimension. In
other words, the electron oscillates back and forth. However, if the motion is random, the
electron in the hydrogen atom would drift away from the nucleus. 

It is well known that, given an ensemble of a quantum state with a distribution of initial
configurations, distributed according to the Born rule, that the Born rule is preserved over
time: The distribution of members of the ensemble satisfy the Born rule at all times
\cite{Towler}. If we insist that this rule must also hold for the velocity sum ($\bv$ + $\bu$),
this greatly restricts the possibilities for the direction of velocity $\bu$, given that the
corresponding speed $\vert\bu\vert$ should agree with the kinetic-energy
expectation-value. This problem does not occur if the velocity $\bu$ is associated with some
sort of motion that does not cause electron translation. In that case, except for ``local
motion,'' an electron, in a state described by a real-valued wavefunction, can be considered
static, and the issue of the static electron, from Bohmian mechanics in such cases, is resolved.

This general idea of two sources of kinetic energy agrees with Salesi's arguments
\cite{Salesi}, obtained from approximations of relativistic quantum-mechanics. He assigns the
kinetic energy from the Madelung velocity $\bv$ to the center of mass motion, and the
Bernoullian velocity $\bu$ to internal energy of the electron. The electron is considered as
having a body occupying space of non-zero measure, undergoing motion relative to its center
of mass.  Salesi also presents an argument that the internal energy of the electron is due to
{\it Zitterbewegung}, from rapid oscillation of the electron \cite{Gerritsma}, a model
that might be able to describe electron spin.

However, there is another possibility: The two electron velocities can be considered
perpendicular. Given, as presented below, that the direction of the velocity vector $\bu$ is
completely changed when including relativistic approximations into the picture, and that the
meaning of the velocity formulae derived from relativistic theory are open to interpretation,
perpendicular velocities is not out of the question. For that reason, other possibilities are
considered in the subsections that follow.

It is worth mentioning that Eq.~(\ref{9904}) indicates that perpendicular velocities,
$\bu\cdot\bv = 0$, implies the satisfaction of the Poisson Equation ($\nabla^2\bar{S} =
-\partial\ln\rho$). It is easily proven that that Poisson-equation satisfaction, implies
$\bu\cdot\bv = 0$, if the wavefunction defined by $R = \sqrt{\rho}$ and $S$, satisfies the
Schr\"odinger equation.

\subsection{Bernoullian fluid-velocity field direction and spin \zlabel{1025}}

\alt{P 30/L 4: Reorganized the material by identifying parts as two separate subsubsection for
  both fluid are particle descriptions, and this is also pointed out to the
  reader. Sec.~(\ref{1025}) is for the fluid description and Sec.~(\ref{qq1010}) is for the
  dynamic-particle description.}

\alt{P 30/L 7: Three sentences added to the paragraph to make a point.}

\alt{P 30/L 10: Modified sentence to make it clear.}

\alt{P 30/L 15: Added ``for these states.'' at the end of a sentence.}

In this subsection, for hydrogen $n$s states and a fluid description, we consider two
Bernoullian velocity-fields: One with zero angular momentum from non-relativistic quantum
mechanics, Eq.~(\ref{p2572}), and a well known one \cite{Holland:99,Colin1,Colin2} from
relativistic quantum mechanics, depending on electron spin, for both fluid are particle
descriptions.  Ignoring spin, the velocity choice (\ref{p2572}) seems to be the best match for
nonrelativistic quantum mechanics, given that it appears in a natural way in the momentum
definition~(\ref{5200}).  Also, the magnitude of this velocity choice appears in the kinetic
energy expression of the Bernoullian Eq.~(\ref{p2574}), and in the expression for the
kinetic-energy expectation value~(\ref{p2922}). As for the velocity from relativistic quantum
mechanics, we show that the direction of this velocity choice can be obtained from
non-relativistic quantum mechanics by requiring a continuity-equation satisfaction for the
hydrogen $n$s states. This continuity-equation involving velocity $\bu$ differs, and should not
be confused with, the one from time-dependent quantum-mechanics---involving velocity $\bv$ in
Bohmian mechanics---that implies probability conservation. Since all stationary states have
time independent probability densities, this time-dependent continuity equation remains
satisfied for states considered.

\alt{P30/L 16: Removed unnecessary clause ``, an equation equivalent to a Schr\"odinger equation.''}

For the one-electron systems that satisfy the energy equation~(\ref{p4791b}), both
$\vert\bu_\pm\vert$ and $\rho_m$ are fixed, and, therefore, only the unit vector
$\mathbf{\hat{s}}_i$ of Eq.~(\ref{p4720x}) is at our disposal.  Also, given
$\mathbf{\hat{s}}_i$, we find no grounds for choosing between the two velocity directions that
satisfy $\bu_+ = -\bu_-$.

\subsubsection{Bernoullian velocity choice from a fluid or internal-energy description}

\alt{P 30/L 22: Removed unnecessary paragraph}


\alt{P 30/L 40: Fixed typo ``Schr\"oodiger'' to ``Schr\"odinger.''}

\alt{P 31/L 5:  Changed ``cancels'' to ``cancel.''}

\alt{P 31/L 7: Made sentence clearer.}

{\bf Zero angular momentum with variable mass.}
Since the $n$s wavefunctions of hydrogen are eigenfunctions of $\hat{L}^2$---the operator for
the square of the magnitude of the angular momentum---and the eigenvalues are zero, it is
reasonable to choose a radially directed momentum, so that the angular momentum vector-field is
the zero function. For the $n$s states of hydrogen, this requires the velocity field given by
Eq.~(\ref{p2572}). However, since the flow is also steady, the flow must have local variable
mass, since a steady radial flow requires infinite mass, if mass is conserved.
A generalization of the continuity equation of fluid mechanics---but not related to a Schr\"odinger
equation---involving this velocity $\bu_\pm$ choice, is obtained from (\ref{p2572}):
\begin{equation} \zlabel{p0258}
\nabla\cdot(\bu_{\pm}\rho) = \pm\fc{\hbar}{2m}\nabla^2\rho
= \mp\fc{2}{\hbar}P,
\end{equation}
where $P$ is the pressure.  Hence, the flows do not locally conserve mass, since
$\{ \mrr\in\mathbb{R}^3\vert P(\mrr) = 0\}$ has measure zero.
These steady flows with variable mass are stabilized by a continuous creation and annihilation
of matter.
Over all space, the flows conserve mass, because the sources cancel the sinks, and this follows
because Eq.~(\ref{p0258}) integrated over all space vanishes. Fluid flows of the local
variable-mass velocity-choice (\ref{p2572}) is explored elsewhere, including the investigation
of the vector fields of the hydrogen atom 1s and 2s states \cite{Finley-Arxiv,Finley-Bern}.

\alt{Modified paragraph for clarity and removed unnecessary equation, that was Eq.~(66).}

{\bf Mass conservation.}

\alt{Removed unnecessary ``assuming there is such a unit vector field $\mathbf{\hat{s}}$.''}

Since local variable mass is a non-classical element for a fluid, and we wish to have the minimum
number of such elements, it is reasonable to investigate the consequences of choosing the unit
vector $\mathbf{\hat{s}}$ of the velocity field $\bu_\pm$ from Eq.~(\ref{p4720x}), or some
modification of this definition, so that the resultant classical continuity-equation yields
\emph{local} conservation of mass, where we consider steady flow. For use below, possibilities
for the Bernoullian velocity $\bu$ are considered, where $\bu$ is not necessarily given by
$\bu_\pm$ from Eq.~(\ref{p2572}). Also, sometimes $\bu$, like $\bu_\pm$, will be unsigned,
having both choices, but in such cases the ``$\pm$'' symbol is not used.

\alt{P 31/L 26. Made changes to the paragraph to make it more comprehensible.}

First consider incompressible flow \emph{along each streamline}, but where the mass density
$\rho_m$ is not uniform. In other words, each streamlines has a constant mass density
$\rho_m$. However, since we have $\nabla\rho_m(\mrr)\ne\mathbf{0}$ for points $\mrr \in
\mathbb{R}^3$ almost everywhere \cite{Finley-Arxiv,Finley-Bern}, $\rho_m$ is not, and cannot be
uniform.  Since such streamlines are on a level surface of $\rho_m$, they have directions
$\bu/\vert\bu\vert$ that are perpendicular to $\nabla\rho_m$, giving
\begin{equation} \zlabel{1100}
\nabla\cdot(\bu\rho_m) = \nabla\rho_m\cdot\bu + \rho_m\nabla\cdot\bu = \rho_m\nabla\cdot\bu = \mathbf{0},
\end{equation}
if mass is conserved.  In other words, local conservation of mass for an electron flow is
satisfied if the volumetric dilatation rate $\nabla\cdot\bu$ vanish, i.e., each fluid
elements has a constant volume.  This is the same requirement for the conservation of mass of a
classical incompressible fluid \cite{Munson,Shapiro,Kelly}, but in this case, the fluid is
incompressible only on a streamline, with the mass density changing from one streamline to
another.

\alt{P 31/L 40: Rewrote paragraph and split it into two paragraphs. Removed an unnecessary equation and 
  added an equation for clarity.}

For the hydrogen $n$s states, the directions of $-\nabla\rho_m$ is in the radial
direction~$\mathbf{\hat{r}}$. Therefore, for the streamline not to cross and also be perpendicular
to $\nabla\rho$ to conserve mass, all the streamline must have a common rotational axis. Let
this rotational axis be the $z$-axis.  Hence, for mass conservation, the streamlines must
always be in the $\hat{\boldsymbol{\phi}}$ direction, where $\phi$ the azimuthal angle of
spherical coordinates, since otherwise they would not always be perpendicular to
$\nabla\rho$, giving local variable mass. Hence, for both conservation of mass and the
proper treatment of kinetic energy, from the consideration of the $n$s states of hydrogen, we
obtain the following velocity choice:
\begin{equation} \zlabel{4418c}
  \bu
  = \pm\vert\bu_\pm \vert\,\hat{\boldsymbol{\phi}}, \quad
\hat{\boldsymbol{\phi}} = \fc{\bu_\pm\times\mathbf{\hat{z}}}{\vert\bu_\pm\times\mathbf{\hat{z}}\vert},
\end{equation}
where $\bu_\pm$ is given by Eq.~(\ref{p2572}). Also, definition~(\ref{4418c}) can be
generalized from hydrogen $n$s states to all quantum states, and this expression is a special
case of definitions~(\ref{p4720x}).

  
The above expression~(\ref{4418c}), derived from non-relativistic quantum mechanics, can be
compared with the velocity field formula obtained for spin one-half particles from
approximations of the Pauli equations \cite{Holland:99,Colin1,Colin2,Salesi}:
\begin{equation} \zlabel{4261}
  \bu = \bu_\pm\times\mathbf{\hat{z}} = \fc{\nabla\rho}{m\rho}\times\mathbf{s}_\pm
  = \vert\bu_\pm\vert\sin\theta\,\hat{\boldsymbol{\phi}},
    \qquad \mathbf{s}_\pm = \pm\fc{\hbar}{2}\mathbf{\hat{z}},
\end{equation}
also giving streamlines for the $n$s states of hydrogen with mass conservation, and
it is easily verified that
$\nabla\cdot\bu = \mathbf{0}$ is satisfied using spherical coordinates, giving incompressible
flow. Also, note that this fluid-velocity choice from relativistic quantum-mechanics has the same
direction as the non-relativistic one~(\ref{4418c}) above. However, definition
(\ref{4261}) does not satisfy $\vert\bu\vert = \vert\bu_\pm \vert$, except on the $x$-$y$
plane where $\bu$ and $\mathbf{\hat{z}}$ are perpendicular, and, therefore, velocity choice
(\ref{4261}) does not not satisfy the
Bernoulian equation~(\ref{p2574}), an equation implied by a Schr\"dinger equation.



\alt{Add a new paragraph for additional information for the literature.}

In turns out, however, that the velocity choice (\ref{4261}) and the non-relativistic
one~(\ref{4418c}) with conserved mass are related. Salesi also demonstrated in his treatment,
by using equalities from relativistic quantum mechanics, that the velocity magnitude, or speed,
from (\ref{4261}) reduces to $\vert\bu_\pm\vert$, so choice (\ref{4261}) reduced to
(\ref{4418c}). In addition, he demonstrated that such a velocity choice satisfies a continuity
equation, also from relativistic quantum mechanics. Therefore, the velocity
choice~(\ref{4418c}) seems reasonable, and there are three identifications of its possible source: a
fluid-velocity field, internal energy of a static electron, or as a particle velocity,
considered below.

{\bf Summary with a fluid model.}
The requirements of zero-angular momentum and the agreement with the kinetic-energy
expectation-value is satisfied by the local variable-mass choice (\ref{p2572}). This choice
also satisfies (non local) mass conservation.  A zero angular momentum for a fluid flow over
all space is incompatible with local mass conservation with a finite total mass; mass must be
created in some regions and destroyed in others.  The requirement of local mass-conservation
satisfaction and the agreement with the kinetic-energy expectation-value is satisfied by
choice~(\ref{4418c}), which includes information from relativistic quantum mechanics.

\subsubsection{Velocity Choice with a dynamic particle description \zlabel{qq1010}}

\alt{P 32/L 4:  ``the ``random'' is removed for the second to last sentence.}

For one-body systems, since the fluid streamlines are the possible paths for the particle, it
can be useful to think about the fluid velocity field $\bu_\pm$ when considering the electron
as a particle. If we choose the velocity to be the minus choice of $\bu_\pm$ given by
(\ref{p2572}), then an $n$s electron moves steadily away from the nucleus, suggesting an
unstable atom, and a similar argument is applicable for the plus choice of $\bu_\pm$, where the
electron heads towards the nucleus.  Therefore, some modification, involving addition elements,
is needed to make the system stable. For example, let $\vert\bu_\pm\vert$ be the average speed
of one-dimensional radially-directed motion, discussed above. Another interpretation is that an
electron in the hydrogen atom with zero-spin is unstable, suggesting the need for a
relativistic correction involving spin, as discussed above for the fluid description, which we
consider below for a particle description.
The derivation of the velocity choices (\ref{4418c}) and (\ref{4261}) for a fluid description
is easily modified for a particle description, by replacing mass-conservation satisfaction with
continuity-equation satisfaction. For these choices, the $n$s electron is in a stable circular
orbit centered on the $z$ (vertical) axis with angle $\theta$ off this axis.  Because of the
cross product in (\ref{4261}), the speed, a scalar field, is proportional to $\sin\theta$. With
the requirement that the electron motion yields the correct kinetic energy, in order to satisfy
the energy equation~(\ref{p4791b}), the electron must be restricted to the $x$--$y$ plane,
giving ($\theta = \pi/2$), or else, it will not have sufficient kinetic energy. It is
demonstrated elsewhere \cite{Finley-Arxiv,Finley-Bern} that the 1$s$ electron speed calculated
by Eq.~(\ref{p2572}), which is equal to the speed from choice from (\ref{4418c}), or
(\ref{4261}) restricted to the the $x-y$ plane, is a constant, and it is the same speed as the
one for the first Bohr orbit of hydrogen.  Colijn and Vrscay use velocity choice~(\ref{4261})
to examine the Bohmian trajectories of electronic states of the hydrogen atom \cite{Colin1}.

\alt{Changed ``Conclusion'' to ``Summary.''}

\alt{P33/L 42: Split one sentence into two.}

\alt{P33/L 43: Removed mention of ``random motion.''.}

\alt{P33/L 46: Removed unnecessary phrase.}

\alt{Removed last sentence of paragraph and added a sentence.}

{\bf Summary for $ns$ states of hydrogen for a dynamic particle description.}  The velocity choice
(\ref{p2572}) requires additional elements for stability. One-dimensional motion with the
average speed given by choice (\ref{p2572}) is one possibility.  The requirements of
continuity-equation and kinetic-energy expectation-value satisfaction with velocity
choice~(\ref{4261}) is given by a spin one-half electron with a circular orbit around the
nucleus, on the $x-y$ plane, and perpendicular to the spin vector $\mathbf{s}_\pm$, giving two
possible direction, one for each spin state.  The Bohr model of atomic hydrogen is a special
case with a known, and fixed, orbit radius.  Velocity choice~(\ref{4418c}) satisfies both the
continuity equation and kinetic energy requirements, with no modifications.

\alt{P 24/L 11 added a paragraph to give another possibility.}

The restriction of the hydrogen $n$s states to the $x-y$ plane is not satisfactory, since one
would want the electron to be able to visit any location, in a manner that has some sort of
association with the probability density $\rho$, required for applications with energy
fields spread over all space. The velocity choice~(\ref{4418c}) is not incompatible with this requirement.

\alt{Added a Conclusion.}

{\bf Overall conclusion.} Velocity choice (\ref{p2572}) describes spin-zero particles as fluids
with local, variable mass. For a given stationary state, velocity choice~(\ref{4418c})
describes a spin one-half particle, either as a static particle with some sort of, unknown,
internal energy, or as a particle orbiting, that is not necessarily on a plane containing the
nucleus, having orbital-angular momentum. Since the curl of the velocity field (\ref{4418c})
is, in general, nonzero, if such a particle is endowed a body, it spins while it orbits, and,
therefore, has spin-angular momentum. For a corresponding fluid identification, the flow is
rotational with a forced vortex.

\alt{Added self contained summary and removed outline section after the Introduction.}

\section{Summary \zlabel{4925}}

To reduce clutter, except for two equations near the end, this summary gives the equations for
the special case of one-body.  These become $n$-body equations, using the notational system of
Sec.~\ref{5147}.

A generalized Euler equation of fluid dynamics, displayed below, is derived. This equation is
implied by the $n$-body time-dependent Schr\"odinger equation, where the fields from the
equation have the same domain as the wavefunction. These fields are maps of the probability
distribution ($\rho = \vert\Psi\vert^2$) and the phase factor $S$ of the wavefunction in polar
form~($\Psi=Re^{iS/\hbar}$, $R^2 = \rho$).  For example, one of the velocity fields is ($\bu =
-(\hbar/2)\nabla\ln\rho$). The statements in this summary of the equivalence of equations, and
an implication, require these field definitions as auxiliary constraints. The field definitions
are given near the end of this summary.

The derived Eulerian equation, mentioned above, for the special case of one-body quantum-states,
can be viewed as a sum of two equations~(\ref{8888d}),
($\mathbb{EQ}_\bu + \mathbb{EQ}_\bv$),
corresponding to two interacting systems: One equation $\mathbb{EQ}_\bv$ with conserved mass, involving
velocity $\bv$ and pressure $P_\bv$; one equation $\mathbb{EQ}_\bu$ with variable mass, involving
velocity~$\bu$, pressure $P_\bu$, and the external potential $U$ from the Schr\"odinger
equation. Explicitly, these equations are the following:

\Eqline{\mathbb{EQ}_\bv = \rho_m\partial\bv \; +\; \fc{1}{2}\rho_m\nabla v^2 + \nabla P_\bv, \quad \rho_m = m\rho,\;\; v = \vert\bv\vert,}

\Eqline{\mathbb{EQ}_\bu = \partial(\rho_m\bu) + \fc12 \rho_m\nabla u^2 +
  \nabla\cdot(\rho_m\bu)\bu + \nabla P_\bu + \rho\nabla U,} where $\rho_m$ is the mass density
for the corresponding classical fluid, and $\partial$ is the partial time derivative.

An $n$-body generalization of the Euler equation ($\mathbb{EQ}_\bu + \mathbb{EQ}_\bv$) is also
derived.  In turn, this Eulerian equation is derived from the gradient of a total-energy
equation:

\Eqline{\bE = \fc{1}{2}m\left(u^2 + v^2\right) + \rho^{-1}\left(P_\bu + P_\bv\right) + U,\quad \labb{5830d}}
that is also a generalization of the Bernoulli equation of fluid dynamics, where $U$ is the sum
of the external potential and the two-body electron-electron repulsion-energy operator.  In turn, this
Bernoullian-equation generalization is a sum of two, energy-field equations ($\bE\defi
\bE_S + \bE_\theta$): 

\Eqline{\bE_S = \fc12 m u^2 + \fc{1}{2}m v^2 + P_\bu\rho^{-1} + U, \quad \labb{5830c} }
\Eqline{\bE_\theta = \rho^{-1}P_{\bv}. \quad \labb{5830d}}
Together, these two equations are demonstrated to be equivalent to the time-dependent
Schr\"odinger equations.  In the special case of time-independence, $\bE_S$ is the energy
eigenvalue of the Schr\"odinger equation. If, in addition, the wavefunction is real valued, we
have $\bE_\theta = \mathbf{0}$, $v^2 = \mathbf{0}$, and the surviving equation for the energy field $\bE_S$
reduces to the derived, Bernoullian Eq.~(\ref{p2574}): $(\bE_S = \fc12 m u^2 + P_\bu\rho^{-1} +
U$).

The above two displayed energy-equations, for $\bE_S$ and $\bE_\theta$, are also equivalent to
the two equations of Bohmian-mechanics:

\Eqline{-\partial S = \fc{1}{2}m v^2  + Q  + U,\quad \labb{p5200}}

\Eqline{\partial\rho + \nabla\cdot(\rho\bv). \quad \labb{p4288}}
These two equation are Hamilton-Jacobi and continuity equations, respectively.  The quantum
potential $Q$ is demonstrated to given by Equation~(\ref{p5202}): ($Q = \fc{1}{2}m v^2 +
P_\bu\rho^{-1}$).

The three pairs of fields of momentum $(\bv,\bu)$, pressure $(P_\bv,P_\bu)$, and energy
$(E_S,E_\theta)$ in the energy equations above, are defined by the real and imaginary parts of
functions. A kinetic-energy field is also defined. These functions involve the wavefunction and
either the momentum $\hat{P}$ or energy $\hat{H}$ operators of quantum mechanics. The pair of
particle momentums, $m\bv$ and $m\bu$, called momentae, are defined by

\Eqline{\rho^{-1} \Psi^*\hat{\text{P}}\Psi \defi m\bv + im\bu, \quad 
  \labb{5204}}
giving the formulae

\Eqline{m\bv = \nabla S, \quad  m\bu = -(\hbar/2)\nabla \ln\rho \defi \nabla \theta, \quad
  \labb{5977}\!,\, \labb{5924}}
where $S$ and $\theta$ are the momentae potentials, and where $\theta$ is defined by the above
equation (to within an additive constant).  The kinetic-energy field is defined to be
$\vert\Psi^*\hat{\text{P}}\Psi\vert^2/2m$, giving ($mu^2 + mv^2$)/2. Similarly, the
energy fields are defined by

\Eqline{\rho^{-1}\Psi^*\hat{H}\Psi \defi \bE_S + i\bE_\theta \quad \labb{5828}} where the
time-dependent Schr\"odiger equation ($i\hbar\Psi^*\partial\Psi/\rho =
\rho^{-1}\Psi^*\hat{H}\Psi$) is substituted to obtains formulae involving time derivatives:

\Eqline{\bE_S= -\partial S, \quad\bE_\theta = -\partial\theta. \quad \labb{5830}}
Also, the pressures are defined by

\Eqline{(\hbar/2m)\nabla\cdot\left(\Psi^*\hat{\text{P}}\Psi\right)\; \dot{=} -\!P_\bv + i P_\bu.\quad \labb{4007}}
giving the formulae:

\Eqline{P_\bu = (\hbar/2m)\nabla\cdot\rho_m\bu_-,\quad  P_\bv =   -(\hbar/2m)\nabla\cdot(\rho_m\bv).\quad \labb{4002}}

Other relations are derived from the continuity equations involving the pressures and time derivatives:

\Eqline{P_{\bv} = (\hbar/2)\partial\rho,  \quad \labb{4302n}}

\Eqline{\nabla_i P_{\bv} =-\partial(\rho_m\bu_i),\quad  i \in\{1,\cdots n\}. \quad \labb{5025n}}

\Eqline{-(\hbar/2m) \nabla_i^2 P_{\bv} =  \partial P_{\bu i}, \quad i \in\{1,\cdots n\}, \quad \labb{4291}}
where the last two equations reduce to the special case of one-body by suppressing the $i$ and $j$ subscripts.

Sec.~(\ref{4290}) demonstrates energy conservation for the nonuniform energy fields, $\bE_S$
and $\bE_S$, in cases where the wavefunction is a nondegenerate linear-combination of
eigenfunctions of a time-independent Hamiltonian operator $\hat{H}$, where the nondegenerate
linear-combination is not an eigenfunction of the Hamiltonian operator.  Sec.~(\ref{5382})
gives a thorough discussions on the compatibility of the two velocity fields for vector
addition, and the consideration of many possibilities for modifications of the Bernoullian
velocity $\bu$, to obtain the best match from classical mechanics.


\section{Discussion \zlabel{5921}}

Given the wave-particle duality of quantum mechanics, it is not surprising that the Eulerian
equation~(\ref{8888b}) seem to describe two interaction substates, since both of the concepts
indicate a doubling of the degrees of freedom.  The presence of two substates, justifies the
use of more complicated physical models to describe the electron, even though there are
numerous constraints. However, this produces a new problem of uniqueness, at least for the
identification of a model for the kinetic energy coming from the Bernoullian velocity $\bu$:
There seems to be too many possibilities, especially considering the possibilities explored in
Sec.~(\ref{5382}), and the observation that some of the various properties are compatible. For
an over the top example, the electron has a compressible body that stores energy, and rotates,
translates, and/or ``zitters,'' while being in a medium under pressure or tension, gaining and
losing mass as it goes, and, all the while, diffusing through the medium.  To identify the best
model, comparisons with experiment would be useful.

\section*{Declarations}

The author did not receive support from any organization for the submitted work.  The author
has no relevant financial or non-financial interests to disclose.

\section*{Data availability statement}


All data that support the findings of this study are included within the article.




\appendix

\section{The variable mass Euler equation for irrotational flows \zlabel{p8250}}

In the derivation of the Euler equation, the differential form of the momentum-balance equation
is used that does not hold for systems that do not conserve mass.  Since our systems under
consideration have variable mass, we need a form of the Euler equation that does not use the
continuity equation. It is trivial to derive variable mass momentum-balance equations, one
simply uses a standard derivation \cite{Currie}, but refrain from utilizing the continuity
equation, leading to some extra terms in the working equation.  Here we derive the
variable-mass Euler equation for irrotational flows. I follow the vector-calculus approach and
notation of Kelly \cite{Kelly}.

\alt{P 36/L 24: For clarity, gave explicit definitions for mass density $\rho_m$, charge density and
$q\rho$ and included $\rho = \rho_m/m$.}

We start with the momentum balance equation for a fluid subject to a Coulombic body-force
with force per charge $(-\nabla\Phi)$ with mass density $\rho_m$, charge density $q\rho$
and $\rho = \rho_m/m$:
\begin{equation} \zlabel{3920}
\fc{d}{dt}\int_V \rho_m\bu \; dV = \int_S \mathbf{\sigma}\mathbf{n}\, dS + \int_V q\rho(-\nabla\Phi)\, dV,
\end{equation}
where $\mathbf{\sigma}$ is the stress tensor and $\mathbf{n}$ is the normal unit vector to the
surface $S$, the border of the subspace $V$.  First we work on the left-hand side.  Using the Reynolds'
transport theorem, and the definition,
\[
\int_{V}g\fc{d(dV)}{dt}\; \dot{=} \; \int_{V}g\nabla\cdot \mathbf{u} \; dV,
\]
where $g$ is an arbitrary function, we obtain
\[
  \fc{d}{dt}\int_{V(t)}\rho_m\bu \; dV =
  \int_V\left(\rho_m\fc{d\bu}{dt} + \bu\fc{d\rho_m}{dt}\right)\; dV +  \int_V \rho_m\bu\fc{d(dV)}{dt}
\]
\[  
  = \int_V \left(
  \rho_m\pa{\bu}{t} + \rho_m (\grad \bu)\bu + \left(\pa{\rho_m}{t} + \nabla\rho_m\cdot\bu\right)\bu + (\nabla\cdot\bu)\rho_m\bu
  \right)\; dV,
\]
where $[\grad\bu]_{ij} = \partial u_i/\partial x_j$ and $[(\grad\bu)\bu]_i = (\partial u_i/\partial x_j)u_j$, summed over $j$ .  Hence
\begin{equation} \zlabel{2302}
 \fc{d}{dt}\int_V \rho_m\bu\; dV =
\int_V \left(\rho_m\pa{\bu}{t} + \rho_m (\grad\bu)\bu  + \mathbb{M}\bu \right)\, dV,
\end{equation}
where
\begin{equation} \zlabel{4922}
\mathbb{M}  = \pa{\rho_m}{t} + \nabla\cdot(\rho_m\bu)  = \dot{\rho}_m + (\nabla\cdot\bu)\rho_m,
\end{equation}
and the continuity equation for systems that conserve mass is $\mathbb{M} = \mathbf{0}$.  Note
that the definition above permits the use of a product rule of differentiation on the factors
of the integrand: $\rho_m\times\bu\times dV$, where $dV$ is ``considered'' a factor, even
though the symbol is excluded in some notations for integrals.

For the surface integral of Eq.~(\ref{3920}), we apply the  divergence theorem:
\begin{equation} \zlabel{8372}
  \int_S \mathbf{\sigma}\mathbf{n} \, dS  = \int_V \dive {\mathbf{\sigma}} \, dV.
\end{equation}

Substituting Eq.~(\ref{2302}) and (\ref{8372}) into (\ref{3920}) gives the differential
momentum-balance equation with variable mass:
\begin{equation} \zlabel{5973} 
\rho_m\pa{\bu}{t} + \rho_m (\grad\bu)\bu + \mathbb{M}\bu = \dive \mathbf{\sigma} + q\rho(-\nabla\Phi),
\end{equation}
where we removed the integrations and obtained a true statement, since the equation with the
integrations holds for all subspaces $V$.

Next we consider only inviscid flows. By definition, these satisfy $\mathbf{\sigma} =
-p\mathbf{I}$, where $p$ is the pressure.  Using this equality and a vector identity, the first
term on the right-hand side (\ref{5973}) for inviscid fluids becomes
\[
\dive \mathbf{\sigma} = -\dive(p\mathbf{I}) = -\nabla p.
\]
Next we require $\bu$ to be irrotational, i.e., $\nabla\times \bu = \mathbf{0}$. This permits
the use of the following equality:
\begin{equation} \zlabel{2947}
(\grad\bu)\bu = \fc12\nabla u^2, \quad \text{if} \quad \nabla\times \bu = \mathbf{0}. 
\end{equation}
Substituting the above two equations into (\ref{5973}) we obtained the desired equation:
\begin{equation*} 
\rho_m\pa{\bu}{t} + \fc12\rho_m\nabla u^2 + \left(\pa{\rho_m}{t} + \nabla\cdot(\rho_m\bu)\right)\bu = -\nabla p - q\rho\nabla\Phi,
\end{equation*}
and we also used (\ref{4922}).  This equation can also be written
\begin{equation}
 \zlabel{7288} \pa{}{t}(\rho_m\bu) + \fc12\rho_m\nabla u^2 + \nabla\cdot(\rho_m\bu)\bu + \nabla p + q\rho\nabla\Phi = 0.
\end{equation}
This equation is the variable-mass Euler equation for the special case of irrotational flow.
In other words, Eq.~(\ref{7288}) is applicable to flows that are irrotational, compressible, invsicid, and
variable-mass. We also require the body force to be Coulombic, but, obviously,
$q\rho\nabla\Phi$ can be replaced by $\rho_m F$, where $F$ is a force per mass. In the special
case where the flow is steady, incompressible, and mass is conserved, the division of
(\ref{7288}) by $\rho_m$, followed by integration, yields the Bernoulli equation~(\ref{p2574}).

In the special case where mass is conserved, explicitly given by
\[
\pa{}{t}\rho_m + \nabla\cdot(\rho_m\bu)\bu = \mathbf{0}.
\]
Eq.~(\ref{7288}) reduces to the familiar form:
\begin{equation}
 \zlabel{7288b} \rho_m\pa{\bu}{t} + \fc12\rho_m\nabla u^2 +  \nabla p + q\rho\nabla\Phi = 0.
\end{equation}




\section{Equalities for the Velocities \zlabel{p7722}}

Next we show that
\begin{equation} \zlabel{p2880}
  \bu_{i\pm} = \pm\text{Re}\left(\fc{\hbar}{m}\fc{\nabla_i\Psi}{\Psi}\right),
  \quad \bv_i = \text{Im}\left(\fc{\hbar}{m}\fc{\nabla_i\Psi}{\Psi}\right),
\end{equation}
where the second one is well known.  To reduce clutter we suppress the $i$ subscripts.  We
require $\bu_\pm$ to be defined by (\ref{p4720}) and we use $\mathbf{v} = \nabla S/m$, from
(\ref{p3312}), to define $\mathbf{v}$.  Starting with the ansatz (\ref{p4225}) $\Psi=
Re^{iS/\hbar}$ we have
\[ \nabla\Psi = (\nabla R) e^{iS/\hbar} + i\hbar^{-1}Re^{iS/\hbar}\nabla s,\]
\[ \fc{\hbar}{m}\fc{\nabla\Psi}{\Psi} = \fc{\hbar}{m}(\nabla R)R^{-1} + i\fc{\nabla S}{m}. \]
Taking the imaginary part of this equation, and using $\mathbf{v} = \nabla S/m$, gives the second
one from~(\ref{p2880}). The real part of the above equation is
\[
\pm\text{Re}\left(\fc{\hbar}{m}\fc{\nabla\Psi}{\Psi}\right) = \pm \fc{\hbar}{m}\fc{\nabla R}{R}.
\]
Starting with (\ref{p4720}) for $\bu_\pm$ we have
\[
\bu_{\pm}= \pm\fc{\hbar}{2m}\fc{\nabla\Upsilon}{\Upsilon}
= \pm\fc{\hbar}{2m}\fc{\nabla R^2}{R^2},
= \pm\fc{\hbar}{m}\fc{\nabla R}{R}.
\]
So the first one from (\ref{p2880}) is also true.






\bibliography{ref}

\bibliographystyle{unsrt}
\end{document}